\documentclass[reprint,showpacs,superscriptaddress,english,twocolumn,notitlepage,longbibliography,nofootinbib]{revtex4-1}
\usepackage[utf8]{inputenc}
\usepackage{color}
\usepackage{array}
\usepackage{verbatim}
\usepackage{multirow}
\usepackage{amsmath}
\usepackage{amssymb}
\usepackage{dsfont}
\usepackage{graphicx}
\usepackage{esint}
\usepackage[unicode=true,
 bookmarks=true,bookmarksnumbered=true,bookmarksopen=false,
 breaklinks=true,pdfborder={0 0 0},backref=false,colorlinks=true]
 {hyperref}
\usepackage{enumitem}
\usepackage{titlesec}
\usepackage{cleveref}
\usepackage{longtable}
\usepackage{bbm}
\usepackage[normalem]{ulem}
\usepackage{listings}

\definecolor{mygreen}{RGB}{28,172,0} 
\definecolor{mylilas}{RGB}{170,55,241}

\hypersetup{
citecolor={blue},
urlcolor={magenta}
}

\newcommand{\PreserveBackslash}[1]{\let\temp=\\#1\let\\=\temp}
\newcolumntype{C}[1]{>{\PreserveBackslash\centering}p{#1}}
\newcolumntype{R}[1]{>{\PreserveBackslash\raggedleft}p{#1}}
\newcolumntype{L}[1]{>{\PreserveBackslash\raggedright}p{#1}}

\usepackage{cleveref}
\crefname{appendix}{Appendix}{Appendices}
\crefname{equation}{Eq.}{Eqs.}
\crefname{figure}{Fig.}{Figs.}
\crefname{table}{Table}{Tables}
\crefname{section}{Section}{Sections}

\renewcommand{\paragraph}[1]{\vspace{0.2cm}{\bf \textit{#1}}}
\def\ie{{\it i.e.},\ }
\def\eg{{\it e.g.},\ }

\newcommand{\mbb}{\mathbb}

\newcommand{\td}{\widetilde}

\def\pare#1{\left( #1 \right)}

\def\abs#1{\left| #1 \right|}

\def\nono{\nonumber}

\def\pr{\prime}

\def\hx{\hat{x}}
\def\hy{\hat{y}}
\def\hz{\hat{z}}

\def\kk{\mathbf{k}}

\begin{document}
\title{Delocalization Transition of Disordered Axion Insulator}
\author{Zhi-Da Song}
\email{zhidas@princeton.edu}
\author{Biao Lian}
\affiliation{Department of Physics, Princeton University, Princeton, New Jersey 08544, USA}
\author{Raquel Queiroz}
\affiliation{Department of Condensed Matter Physics, Weizmann Institute of Science, Rehovot 7610001, Israel}
\author{Roni Ilan}
\affiliation{Raymond and Beverly Sackler School of Physics and Astronomy, Tel Aviv University, Tel Aviv 69978, Israel}
\author{B. Andrei Bernevig}
\affiliation{Department of Physics, Princeton University, Princeton, New Jersey 08544, USA}
\affiliation{Physics Department, Freie Universitat Berlin, Arnimallee 14, 14195 Berlin, Germany}
\affiliation{Max Planck Institute of Microstructure Physics, 06120 Halle, Germany}
\author{Ady Stern}
\email{adiel.stern@weizmann.ac.il}
\affiliation{Department of Condensed Matter Physics, Weizmann Institute of Science, Rehovot 7610001, Israel}
\date{\today}

\begin{abstract}
The axion insulator is a higher-order topological insulator protected by inversion symmetry.
We show that under quenched disorder respecting inversion symmetry {\it on average}, the topology of the axion insulator stays robust, and an intermediate metallic phase in which states are delocalized is unavoidable at the transition from an axion insulator to a trivial insulator.
We derive this conclusion from general arguments, from classical percolation theory, and from the numerical study of a 3D quantum network model simulating a disordered axion insulator through a layer construction.
We find the localization length critical exponent near the delocalization transition to be $\nu=1.42\pm 0.12$.
We further show that this delocalization transition is stable even to  weak breaking of the average inversion symmetry, up to a critical strength.
We also quantitatively map our quantum network model to an effective Hamiltonian and we find its low energy k$\cdot$p expansion.
\end{abstract}

\maketitle

\textit{\textbf{Introduction}}
Localization of electronic states in disordered systems has been extensively studied in the past decades \cite{Anderson1958,abrahams50}.
The existence and characteristics of metal-insulator Anderson transitions is usually determined by the dimension of the system, the symmetries it respects \cite{Zirnbauer1996,Altland1997}, and the topological classification that they lead to.
In particular, studies on the quantum Hall states reveal a profound relation between delocalization and the topology of the electronic state \cite{khmel1983quantization,Pruisken1983,chalker_percolation_1988,Ludwig1994QH}.
Hence an interesting question is how the localization interplays with the full range of band topologies in the past two decades.
For topological insulators protected by nonspatial symmetries \cite{Kane2005Z2,Bernevig2006BHZ,Konig2007HgTe,hasan2010review,qi2011review,kitaev2009periodic,Ryu2010tenfold},  it has been shown that the gapless boundary states are stable against symmetry-respecting disorder \cite{Ryu2010tenfold,Lu2011localization,He2011localization,Chen2011localization,Wang2012localization}, and the phase transition point between phases of different bulk topological numbers has protected extended bulk states at the chemical potential \cite{Ludwig1994QH,fulga_statistical_2014,morimoto_anderson_2015}.
Topological states protected by translation \cite{ringel_strong_2012,Moore2012} or mirror \cite{fu_topology_2012,fulga_statistical_2014}  symmetries are shown to have stable gapless surface states if the crystalline symmetries are respected on average by the disorder.
However, such analyses do not explore the effect of disorder on bulk states, and do not generalize to the topological states protected by generic crystalline symmetries \cite{fu_topological_2011,mong_antiferromagnetic_2010,Huse2011Inversion,slager2013space,Liu2014TCI,Fang2015TCI}, such as higher-order topological insulators \cite{ZhangFan2013Axion,benalcazar_quantized_2017,benalcazar_electric_2017,schindler_higher-order_2018,schindler2018higher,langbehn_reflection-symmetric_2017,song_d-2-dimensional_2017,Ezawa2018Higher}.
Very recently, some numerical studies have shown the robustness of the higher-order topological insulators \cite{Su2019disorder,Araki2019disorder,wang2020disorder,Li2020Quadrupole}, but an understanding of this robustness and of the delocalization transitions of these insulators is still lacking.

\begin{figure}[t]
\centering
\includegraphics[width=1\linewidth]{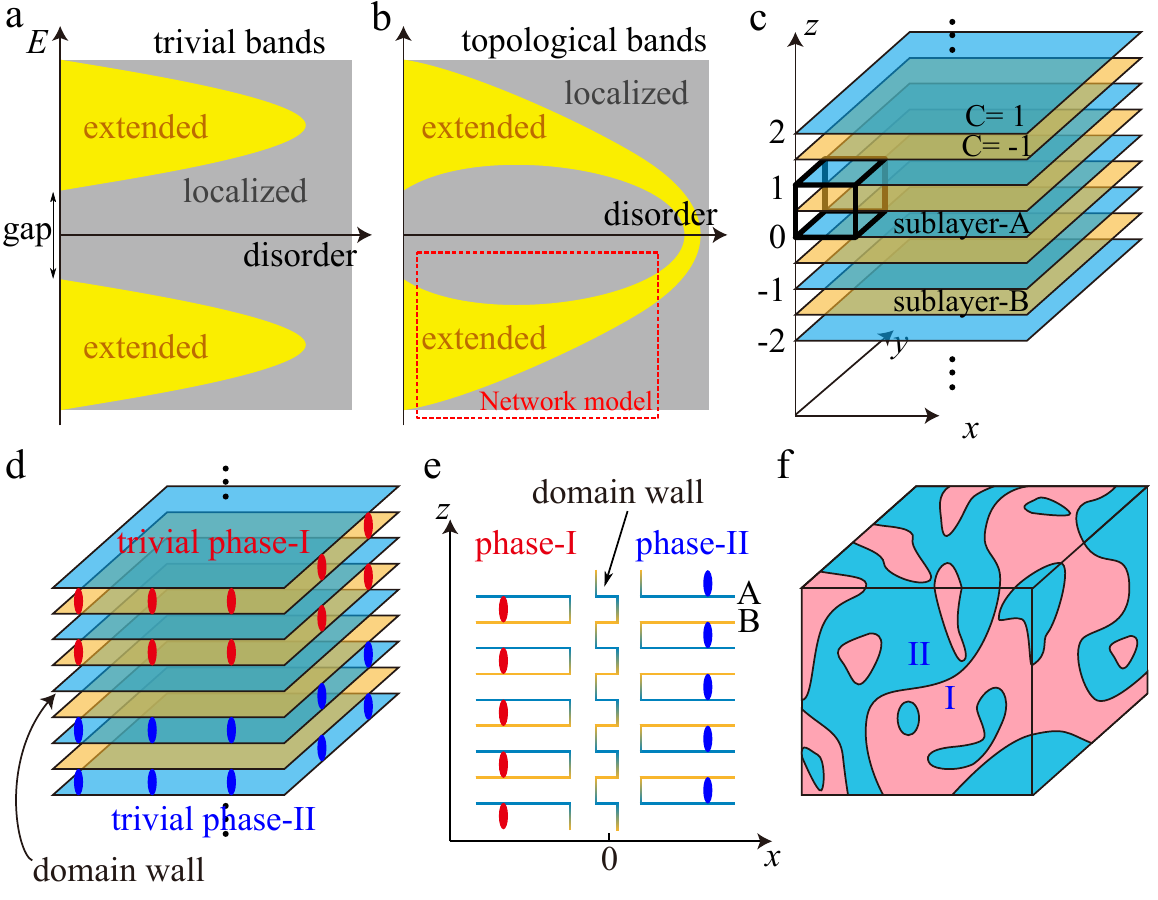}
\caption{ Localization and band topology.
(a-b) Localized (grey) and delocalized (yellow) regions in the spectrum as a function of disorder for 3D trivial and axion insulators, respectively.
(c) Layer construction for the axion insulator, where the black box denotes the 3D unit cell.
Sublayers $A$ (blue) and $B$ (orang) are decorated by 2D Chern insulators with Chern numbers $C=1,-1$, respectively.
Each unit cell has eight inversion centers $(t_x/2,t_y,t_z)$ ($t_{x,y,z}=0,1$), all of which lie in a Chern layer.
(d) Two possible inversion-breaking dimerized phases (I and II), which are inversion partners.
(e) Side view of the domain wall in the $xz$ direction.
(f) A disordered axion insulator with random dimerizations, where the red and blue regions represent phases I and II, respectively.
The domain walls between phases I and II are Chern layers and host extended states.
\label{fig:LC}
}
\end{figure}

In a generic three dimensional (3D) electron band, for weak disorder mobility edges develop, dividing the electron states into localized states near the band edges and extended states in the middle of the band (\cref{fig:LC}a).
When the disorder energy scale exceeds the bandwidth, a topologically trivial band has \emph{all} the states localized.
In contrast, for weak disorder a two dimensional (2D) single Chern band with Chern number $C=1$ has one energy at which states are delocalized. At strong disorder the band is trivialized  by the disorder mixing it with another band of opposite Chern number \cite{khmel1983quantization,Pruisken1983,chalker_percolation_1988,Ludwig1994QH,wang_universal_2014} (\cref{fig:LC}b).
The transition between a $C=1$ state to a trivial $C=0$ state occurs through the occurrence of delocalized states, this time in the gap that separates the two bands.
All the topological insulating phases in 2D and 3D protected by local symmetries (\eg time-reversal) are believed to have similar delocalization transitions, although for weak disorder in 3D, as well as in nonmagnetic 2D systems with spin-orbit coupling, there presumably is a non-zero range of energies with delocalized states.

In this work, we examine whether disorder induce transitions that are associated with bulk delocalization also for topological crystalline insulators. We focus  on the case of a disordered axion insulator \cite{Vanderbilt2009axion,li2010dynamical,Huse2011Inversion,Ashvin2012Axion}, which has a quantized magneto-electric response (with axion field $\theta=\pi$) and recently identified as a higher-order topological insulator protected by inversion symmetry \cite{ZhangFan2013Axion,Vanderbilt2018Axion,wieder2018axion,Xu2019EuIn2As2,yue2019axion,Zhang2019MnBi2Te4}.
We show that a 3D delocalized metallic phase necessarily arises during the transition from an axion insulator to a trivial insulator as long as the inversion symmetry is respected (or broken weakly enough) on average.
Such a delocalization transition manifests the robustness of the axion insulator topology against disorder.

\textit{\textbf{Layer construction argument}}
We use a layer construction
\cite{song_quantitative_2018,song_topological_2019,Song2017,Huang2017} to argue for the existence of the delocalization transition in a disordered axion insulator.
We consider a 3D crystal with inversion symmetry that maps $(x,y,z)\to(-x,-y,-z)$ and translation symmetry that maps $(x,y,z)\to (x+t_x,y+t_y,z+t_z)$, with $t_{x,y,z}\in\mathbb{Z}$ (\cref{fig:LC}c).
We set the lattice constant as 1.
A shifted inversion operation centered at $(t_x/2,t_y/2,t_z/2)$ consists of the combination of inversion and translation.
There are eight shifted inversion centers in each unit cell, corresponding to $t_{x,y,z}=\{0,1\}$, respectively.
Ref. \cite{MTQC} shows that the axion insulator state can be constructed from weakly coupled Chern insulators sublayers occupying the inversion centers, where for the $A$ sublayers at $z=0,\pm1\cdots$ the  Chern number is $C=1$ and for the $B$ sublayers at $z=\pm\frac12,\pm\frac32\cdots$ it is  $C=-1$ (\cref{fig:LC}c).
The net Chern number in each unit cell is zero.
The topology of the axion insulator relies on the fact that one cannot trivialize the construction without breaking inversion symmetry.
For example, dimerizing each sublayer $A$ at $z\in\mathbb{Z}$ with the sublayer $B$ at either $z+\frac12$ or $z-\frac12$ leads to a trivial insulator, but breaks the inversion symmetry (\cref{fig:LC}d).

Our analysis starts from two dimensions. We consider a slab made of a finite odd number of layers $N_z\gg 1$ and a very large number of unit cells in the $x,y$ directions, $N_{x,y}\gg N_z$.  Topologically the slab is a 2D Chern insulator, say of $C=1$. 
For the construction in \cref{fig:LC}c, the system carries a chiral edge mode moving along the $x-z$ and $y-z$ sides, while inversion symmetry relates the position of the edge modes on two opposite surfaces. 
However, neither topology nor symmetry fully determine the spatial distribution of the chiral mode in the $z$--direction, which may be confined to a small region or spread over the entire sides.
Furthermore, the chiral edge mode may be moved up and down in the $z$-direction by applying local inter-layer couplings limited to the sides where it flows, without affecting the bulk.

Being a $C=1$ Chern insulator, weak disorder localizes all bulk states except states close to two critical energies $E_{c,1}, E_{c,2}$, one per band. 
(We assume each layer is a two band model for simplicity.)
The system's Chern number is $C=1$ when its chemical potential is between $E_{c,1}$ and $E_{c,2}$, and zero otherwise. It is expected that for a large $N_{x,y}$, $\Delta E$, the width of the energy window around $E_{c,1}$ or $E_{c,2}$ where delocalized states occur, diminishes with increasing system size \cite{Pruisken1983,chalker_percolation_1988}.  
For the Chern number to change, states at the chemical potential must be delocalized, such that they couple edge modes on opposite sides of the system and allow them to mutually annihilate. Assuming that the disorder is uniformly distributed within the system, we conclude that the delocalized bulk states are delocalized in all three dimensions in the slab with $1\ll N_z\ll N_{x,y}$.  
As disorder gets stronger,  $E_{c,1}$ and $E_{c,2}$ get closer to one another, until at some critical disorder they become equal, and the system turns trivial at all energies. 
The picture changes when inversion symmetry is broken.  For example, if the two layers within each unit cell are strongly coupled to one another, they become a trivial insulator, in which disorder localizes all states. 
Then, even for a uniformly distributed disorder, the entire transition between $C=1$ and $C=0$ happens at a single unpaired layer, and the delocalized states occur only in that layer, typically away from any inversion plane.

Now we approach the 3D limit, making $N_x,N_y,N_z$ all very large and comparable to one another, while preserving inversion symmetry. As long as $N_z$ is odd, the Chern number $C=1$ when the chemical potential is tuned properly, and there still is a chiral gapless mode encircling the sample on the side surfaces. 
We expect $\Delta E$ to depend on $N_z$, in such a way that in the 3D limit $\Delta E$ stays non-zero, since for 3D (topological or not) systems there is a metallic phase at weak disorder.
Then the critical energies $E_{c,1}$ and $E_{c,2}$ develop into two energy regions of extended states, as shown in \cref{fig:LC}b.
For stronger disorder, in a generic 3D system we expect the ranges of delocalized energies to shrink, until they disappear and all states become localized (\cref{fig:LC}a-b).
Here, however, the non-vanishing Chern number requires the existence of critical disorder at which states in the gap are delocalized.
While this analysis is based on the Chern number that the system carries for an odd $N_z$, the thermodynamic 3D limit should not depend on the parity of $N_z$. 
Adding an additional $C=-1$ layer to the system will not change the localization properties, because the extra layer applies a local perturbation, while the delocalized states are extensive.
Thus, the delocalized states occurring at the band gap at a critical disorder strength will remain even in the absence of a Chern number, and will signal and signify the transition from an axion to a trivial insulator.

The physical picture behind the 3D delocalized states may be understood in the following way.
We first define two types of inversion-breaking dimerized phases (\cref{fig:LC}d): (I) the dimerization neutralizing sublayer $A$ at $z\in \mathbb{Z}$ with sublayer $B$ at $z-\frac12$, and (II) the dimerization neutralizing sublayer $A$ at $z\in \mathbb{Z}$ with sublayer $B$ at $z+\frac12$.
Phase-I and phase-II are inversion partners of each other, and the domain wall between them is a Chern insulator layer.
Note that the domain does not have to be perpendicular to $z$-direction.
For example, as shown in \cref{fig:LC}e,  if the $x<0$ and $x>0$ regions form phase-I and phase-II, respectively, a domain wall with Chern number 1 must exist in the middle (\cref{fig:LC}e). 
We can think of \cref{fig:LC}e as a surface in the $xz$ direction, where blue and orange lines represent chiral states moving right and left, respectively. 
In trivial phase-I (II), the right movers are reflected to (from) the left movers below (above) it.
The mismatch of the flowing implies a vertical chiral state moving down in the middle, which implies a Chern domain wall in the bulk. 
Inversion-breaking disorders can then be simulated by placing random dimerizations in the 3D bulk, so that the bulk randomly forms phase-I and phase-II in different regions (\cref{fig:LC}e).
When the volume fractions of phase-I and phase-II are equal, we say  inversion symmetry is respected on average.
We have only considered the dimerization disorder for simplicity.
As discussed in Ref. \cite{SUP}, considering more complicated disorder configurations will not change the conclusion.


Since each domain wall hosts a 2D Chern insulator with $C=\pm1$, it must host 2D delocalized states at the energy of a delocalization transition.
If the domain walls form an infinitely large cluster, its  extended states become extended states in the 3D bulk.
Then, when the chemical potential enters the region of the energies of these extended states, a 3D delocalization transition would happen, after which the system enters the disordered trivial insulator phase. 
On the contrary, if all the domain walls do not extend to infinity, the disordered axion insulator and trivial insulator would be connected without phase transition.
By the classical 3D continuum percolation theory \cite{isichenko_percolation_1992}, the domain walls extend to infinity if the volume fraction $p_1$ of phase-I (or $p_2=1-p_1$ of phase-II) is between 0.17 and 0.83.
Therefore, we expect the 3D delocalization transition to exist if the inversion symmetry is on average respected ($p_1=0.5$) or broken weakly enough ($0.17<p_1<0.83$).

\begin{figure}[t]
\centering
\includegraphics[width=1\linewidth]{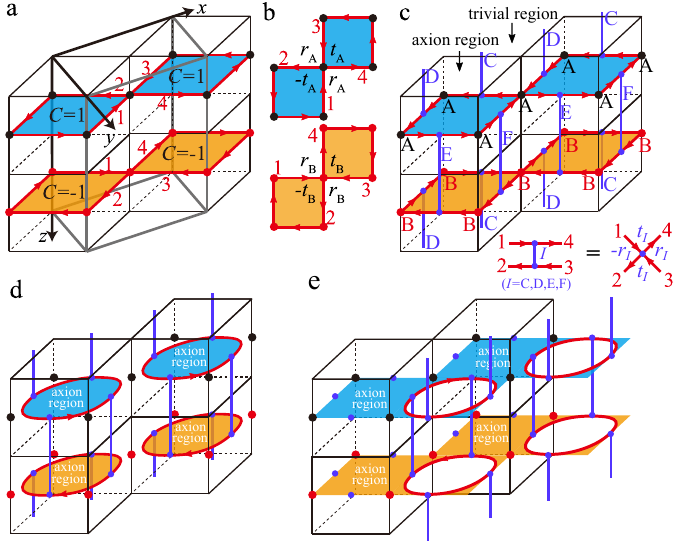}
\caption{ The quantum network model for the axion insulator.
(a) A side view of the 3D system. The blue (orange) regions have a Chern number 1 (-1).
The grey box represents the repeating unit.
The inversion centers are at $(t_x/2,t_y/2,\frac14 + t_z/2)$ for $t_{x,y,z}=0,1$.
The red lines with arrows are the chiral modes surrounding the Chern regions.
(b) Scatterings at the single-layer level. Here $t_{A,B}$ and $r_{A,B}$ are the transmission and reflection amplitudes of the scattering, respectively.
(c) Introducing inter-layer scatterings. 
The nodes $C,D$ ($E,F$) scatters the edge states in the blue layer to the edge states in the orange layer in the above (below).
(d-e) The localized edge states of the Chern regions in the trivial ($t_{A,B}=1$) and axion insulator ($t_{A,B}=0$) limits, respectively.
}
\label{fig:Network3D}
\end{figure}

\textit{\textbf{Quantum network model}}
Our classical percolation argument  neglects quantum tunneling between neighboring domain walls.
To verify the existence of delocalization transition, we study a disordered 3D quantum network model for the layer-constructed axion insulator, which describes Anderson transition with respect to changing chemical potential. The model, which includes only one band for each layer, is suitable for a transition taking place within that band (\cref{fig:LC}b). 
Its analysis also demonstrates the effect of inversion symmetry breaking on this transition.

In the decoupled layers limit,  each sublayer forms a 2D Chalker-Coddington quantum Hall network model \cite{chalker_percolation_1988} (\cref{fig:Network3D}a-b).
For convenience, here we shift the inversion centers to $(t_x/2,t_y/2,\frac14 + t_z/2)$ ($t_{x,y,z}=0,1$) such that the Chern layers are in the $z=\frac14$ and $z=\frac34$ planes.
The blue (orange) and empty regions in sublayer $A$ ($B$) have $C=1$ ($C=-1$) and $C=0$, respectively, while the red lines represent the chiral edge modes. The amplitude $\psi_i$ of a chiral mode propagating through a bond $i$ gains a (quenched) random propagation phase $e^{i\phi_i}$.
Two chiral modes are coupled by tunneling at the crossings of the red lines.
As shown in \cref{fig:Network3D}b, the two outgoing modes ($\psi_2,\psi_4$) are scattered from  the two incoming modes ($\psi_1,\psi_3$) as
\begin{equation}
\psi_2 =-t_{A,B} \psi_1 + r_{A,B} \psi_3,\qquad
\psi_4 = r_{A,B} \psi_1 + t_{A,B} \psi_3, \label{eq:scater}
\end{equation}
where $t_{A,B}=\cos\theta_{A,B}$ and $r_{A,B}=\sin\theta_{A,B}$ are referred to as the transmission and reflection amplitudes in sublayer $A$ and $B$, respectively, which we assume are spatially uniform.
We choose $t_{A,B}$ and $r_{A,B}$ as real numbers because we can absorb their phases into the propagating phases $\phi_i$. 
In the trivial phase of the layers $t_{A,B}=1$ ($\theta_{A,B}=0$). Then, the chiral modes form local loops surrounding the $C=\pm1$ regions and we can continuously shrink the $C=\pm1$ regions to zero.
A non-trivial Chern phase forms when $t_{A,B}=0$ ($\theta_{A,B}=\frac{\pi}2$), where the trivial regions can be shrunk to zero.
Thus tuning $\theta$ from $\frac{\pi}2$ to $0$ simulates tuning the chemical potential from below to above the Chern band.
At the single energy $\theta_{A,B}=\frac{\pi}4$, states in each layer are delocalized. The sublayers go through a phase transition from $C=\pm1$ at $\frac{\pi}4<\theta_{A,B}\le\frac{\pi}2$ to $C=0$ at $0\le\theta_{A,B}<\frac{\pi}4$ \cite{chalker_percolation_1988}.

The decoupled layers limit is inversion symmetric without disorder, \ie with spatially uniform propagation phases $\phi_i$.
Looking at the system as 3D, the pillars (\cref{fig:Network3D}c) containing the colored regions of sublayers A or B can be thought as regions of axion insulators, because each of them has a single Chern layer passing through the inversion center.
The complementary empty regions can be thought as trivial insulator regions.
We emphasize that there is no explicit relation between the axion or trivial regions and the phase-I or phase-II shown in \cref{fig:LC}. 
Both the axion regions and trivial regions are centrosymmetric by themselves, while phase-I and phase-II transform to each other under the inversion.
Turning on the disorder (randomness in phases $\phi_i$) breaks inversion symmetry, but preserves it on average when the $\phi_i$ are uniformly random everywhere.

We introduce inter-layer scattering nodes at the mid-points of each square, half way between the the intra-layer ones, represented by blue vertical lines in \cref{fig:Network3D}c-e. On each square there are four scattering nodes. Nodes of the $C,D$ types couple blue layer edge modes to the orange layer edge modes in the layer above, while $E,F$ types couple the blue layer edge modes to the layer below. Going counter-clockwise along the square, the order of nodes is $C,D,E,F$. 
We parametrize the transmission and reflection amplitudes in the nodes  $t_{I}=\cos\theta_{I}$ and $r_{I}=\sin\theta_{I}$ ($I=C,D,E,F$), respectively.
More details of the scattering parameters are given in Fig. S1 in Ref. \cite{SUP}.
We use four variables $\mu,\gamma,\eta,\delta$ to parameterize the angles:
\begin{equation}
\theta_A= \frac{\pi}{4} + \mu - \eta,\qquad
\theta_B= \frac{\pi}{4} + \mu + \eta,\label{eq:theta-AB}
\end{equation}
\begin{equation}
\theta_C=\theta_D= \gamma(1 - \delta),\qquad
\theta_E=\theta_F=\gamma(1-\delta) + \delta\frac{\pi}2,\label{eq:theta-CE}
\end{equation}
$\mu$ can be interpreted as the chemical potential, $\eta$ tunes the potential energy difference between two sublayers, $\gamma$ and $\delta$ determine the inter-layer couplings.
Inversion transforms the nodes $C,D$ to $E,F$, respectively (\cref{fig:Network3D}), and therefore inversion symmetry is broken on average when $\delta$ is non-zero.
We set $\gamma=\pi/8$ in the rest of this work such that the inter-layer coupling is weak compared to the intra-layer couplings.
As explained in the following paragraphs, the insulating limits are independent with $\gamma$, hence the choice of $\gamma$ does not qualitatively change the phase diagram of the quantum network model. 

We now study the delocalization transitions with respect to the chemical potential ($\mu$), the potential difference between two layers ($\eta$), and the inversion symmetry breaking ($\delta$).
For an inversion symmetric (on-average) system $\eta=\delta=0$.
The sublayers are either both trivial or both topological.
When $\mu=-\frac{\pi}4$, one has $t_{A,B}=1$, and the chiral modes surrounding the $C=\pm1$ regions are closed in each layer and but are vertically connected to the closed chiral modes in the nearby layers (\cref{fig:Network3D}d).
The axion regions can then be adiabatically shrank to zero, so the 3D bulk is in the trivial insulator phase.
When $\mu=\frac{\pi}4$, the chiral modes flow surrounding the trivial regions ($t_{A,B}=0$) as shown in \cref{fig:Network3D}e, so the trivial regions can be shrank to zero, and the 3D bulk is in the axion insulator phase.
In this case, each Chern layer contributes to a chiral mode on the side surface of the system.
Therefore, tuning $\mu$ from $-\frac{\pi}4$ to $\frac{\pi}4$ tunes the chemical potential from the bottom to the top of the topological bands of the axion insulator (\cref{fig:LC}b). In particular, when $\mu=0$, $\theta_{A,B}$ are equal to $\pi/4$, and the 3D bulk must be delocalized because the chiral modes form a connected network, corresponding to the region of delocalized states in \cref{fig:LC}b. 

In contrast, varying $\eta$ from 0 to $\pi/4$ for $\mu=\delta=0$, each sublayer $A$ becomes a trivial insulator ($\theta_{A}=0$), while each sublayer $B$ becomes a Chern insulator with $C=-1$ ($\theta_{B}=\frac{\pi}2$).
Therefore, $\eta$ drives the system into a 3D QAH insulator.

In the end we consider strong inversion breaking (on average).
When $\delta=1$, there is $t_C=t_D=1$, $r_C=r_D=0$, $t_E=t_F=0$, $r_E=r_F=1$, and hence a blue layer is decoupled from the orange layer above it but is fully coupled to the orange layer below it.
The 3D network decomposes into disconnected 2D slices in the $z$-direction.
Since each slice has a vanishing Chern number, there is no guaranteed delocalized state.
Therefore, no delocalization transition with respect to $\mu$ is expected if $\delta=1$.
See Ref. \cite{SUP} for more details.

\begin{figure}[t]
\centering
\includegraphics[width=1\linewidth]{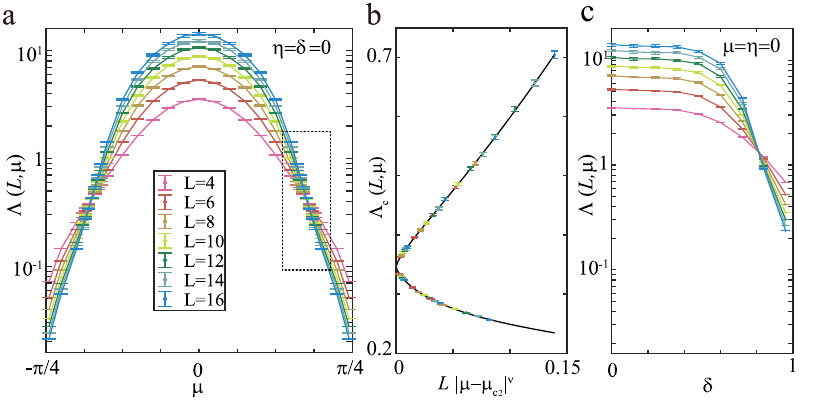}
\caption{Numerical results. (a) The normalized localization length $\Lambda$ of the quasi-1D system is plotted as a function of $\mu$ at different system sizes (widths) $L$.
The system is delocalized for $\mu$ between the two Anderson transition points $\mu_{c}\approx\pm 0.56$.
(b) shows the one-parameter scaling of the relevant part of $\Lambda$ around $\mu\approx 0.56$.
The two branches correspond to $\mu>0.56$ and $\mu<0.56$, respectively.
(c) shows the localization transition of $\Lambda$ due to the inversion symmetry breaking on average, where $\delta$ tunes the symmetry breaking strength.
\label{fig:data}}
\end{figure}

\textit{\textbf{Numerical results}}
The localization length of the network model can be computed with a quasi-1D geometry \cite{mackinnon_one-parameter_1981,chalker_percolation_1988, mackinnon1983scaling}. Technical details are given in Ref. \cite{SUP}. 
A quasi-1D system is always localized, with the localization length depending on the transverse dimension $L$. 
The object of interest is the normalized localization length $\Lambda = \lambda/L$ \cite{mackinnon_one-parameter_1981,mackinnon1983scaling}. 
When $\Lambda$ is finite or divergent in the $L\rightarrow\infty$ limit, the 3D states are delocalized. 

We first focus on the case where inversion symmetry is satisfied on average, \ie $\delta=0$.
($\delta$ is defined in \cref{eq:theta-CE}.)
For $\eta=0$, \cref{fig:data}a shows $\Lambda(\mu,L)$ as a function of $\mu$ and $L$.
At $\mu=0$, $\Lambda(\mu,L)$ increases with $L$, which implies 3D delocalized states.
In contrast, at $\mu=\pm\frac{\pi}4$, $\Lambda(\mu,L)$ decreases with $L$ and approaches zero as $L\to \infty$, implying localized states.
As we discussed earlier in \cref{fig:Network3D}, $\mu=-\frac{\pi}4$ and $\mu=\frac{\pi}4$ correspond to the trivial insulator and axion insulator phases, respectively.
\cref{fig:data}a indicates that there is a delocalized metallic phase between them with the two delocalization Anderson transitions happening at $\mu_{c}\approx \pm 0.56$, where $\Lambda(\mu,L)$'s for different $L$'s cross each other.

On the insulator side of the transitions, the 3D localization length diverges as $\xi \sim |\mu-\mu_{c}|^{-\nu}$, with a universal exponent $\nu>0$.
For sufficiently large $L$, $\Lambda(\mu,L)$ is subject to the one-parameter scaling of the single parameter $L/\xi$ \cite{mackinnon_one-parameter_1981,mackinnon1983scaling}. When $L$ is small, $\Lambda(\mu,L)$ also contains $L$ dependent irrelevant terms because of the finite-size effect, and
assumes the following form \cite{slevin_corrections_1999}:
\begin{equation}
\Lambda(\mu,L)=G_0((\mu-\mu_c)L^{\frac1{\nu}}) + L^{y} G_1((\mu-\mu_c)L^{\frac1{\nu}}). \label{eq:Lambda-fit}
\end{equation}
Here $y<0$ is an irrelevant scaling exponent, and $G_{i}(x)$ ($i=0,1$) are undetermined functions which we keep up to the third order $G_{i}(x) = g^{(i)}_0 + g^{(i)}_1 x + g^{(i)}_2 x^2 + g^{(i)}_3 x^3$.
We fit the parameters by the least square method \cite{SUP} for the data points in the dashed rectangular in \cref{fig:data}a.
\cref{fig:data}b shows the relevant part $\Lambda_c=G_0$ as a function of $L|\mu-\mu_c|^\nu$.
The universal exponent from our fitting is $\nu=1.42\pm0.12$, which is close to that of the 3D Anderson transition under magnetic field (where $\nu$ is found $1.3\pm0.15$ \cite{henneke_anderson_1994}, $1.45\pm0.25$ \cite{chalker_3D_1995}, $1.43\pm0.04$ \cite{Slevin1997-exponent}, and $1.443\pm0.006$ \cite{Slevin2016-exponent}).

\begin{figure}[t]
\centering
\includegraphics[width=1\linewidth]{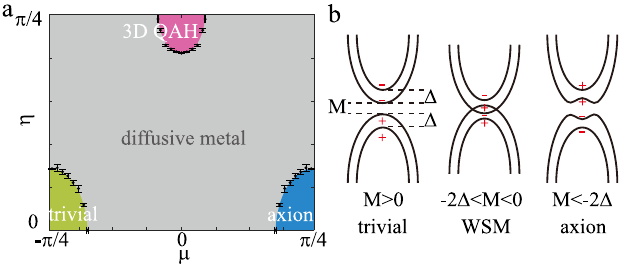}
\caption{
Disordered topological phases.
(a) Phase diagram in the parameter space of $\mu$ and $\eta$. $\delta$ is set to zero.
(b) Gap closing transition from trivial insulator to axion insulator.
The $\pm$ symbols represent the parities of the Bloch states.
\label{fig:band}}
\end{figure}

We have theoretically presented arguments that strong inversion symmetry breaking leads to localization and showed that in the network model $\delta=1$ corresponds to an inversion-broken localized limit.
By tuning $\delta$ in the metal phase at $\mu=\eta=0$, we observe an Anderson transition at $\delta\approx 0.81$ to the inversion-broken localized phase (\cref{fig:data}c).

Keeping $\delta=0$ and applying the finite-size scaling method to nonzero $\eta$, which represents the potential energy difference between sublayers A and B, we obtain a phase diagram of \cref{fig:band}a in the parameter space of $\mu,\eta$ with inversion symmetry respected on average.
A new insulating phase arises near $\mu=0$, $\eta=\frac{\pi}4$.
For a clean system, at $\mu=0$, $\eta=\frac{\pi}4$, sublayer $A$ is at a $C=0$ state and sublayer B at  $C=-1$, hence this phase is a 3D QAH insulator \cite{yu_quantized_2010}.

\textit{\textbf{Discussion}}
So far we used the chemical potential as the phase transition tuning parameter (\cref{fig:LC}).
The same delocalization transitions can be tuned by other parameters, \eg band gap, for which the transitions happen at gap closings that change the topology of the bands.
\cref{fig:band}b illustrates the gap-closing transition from trivial insulator to axion insulator driven by the inversion of band gap $M$ (see Ref. \cite{SUP} for details).
For $\Delta>0$, the transition involves an intermediate Weyl semimetal (WSM) phase \cite{burkov_weyl_2011,wan_topological_2011,weng_weyl_2015,Lv2013Weyl,xu_discovery_2015}.
In Ref. \cite{SUP}, we quantitatively map the clean quantum network model to an effective Hamiltonian, where the parameter $\mu$ plays the role of $M$.
Therefore, the diffusive metal in \cref{fig:band}a is mapped to the disordered WSM, which was found diffusive \cite{Kobayashi2014Weyl,Huse2014,DasSarma2016WSM}. 
We also show that, with a strong inversion breaking ($\delta=0.85$), the model become fully localized for $-\pi/4\le \mu \le \pi/4$. 

For $\Delta=0$, which can be guaranteed by certain extra symmetries \cite{SUP}, the transition is sharp without a WSM phase.
Our theoretical arguments shows the transition always expand into a diffusive metal phase under disorder that respect the inversion symmetry on average.

We expect the delocalization transitions to be studied in the recently proposed axion insulator materials \cite{mogi2017axion,xiao2018axion,yue2019axion,Xu2019EuIn2As2,li2019intrinsic,Zhang2019MnBi2Te4,Gong_2019,Deng895,zhang2019Mobius} in the future.

\textit{Note added.}
We are aware of a related work \cite{li2020critical} focusing more on the surface delocalization transition. 
Their results, when overlap, are consistent with ours.

\begin{acknowledgments}
We are grateful to Xiao-Yan Xu and Yuanfeng Xu for useful discussions.
B. A. B. and Z.-D. S. were supported by the DOE Grant No.~DE-SC0016239, the Schmidt Fund for Innovative Research, Simons Investigator Grant No.~404513, and the Packard Foundation. Further support was provided by the NSF-EAGER No.~DMR 1643312, NSF-MRSEC No.~DMR-1420541,  and ONR No.~N00014-20-1-2303, Gordon and Betty Moore Foundation through Grant GBMF8685 towards the Princeton theory program.
The development of the network model is supported by DOE Grant No. DE-SC0016239.
R. I. and B. A. B. also acknowledge the support from BSF Israel US foundation No.~2018226.
A. S. acknowledges supports from the European Research Council (Project LEGOTOP), the National Science Foundation under Grant No. NSF PHY-1748958, the Israel Science Foundation - Quantum Program, and the RC/Transregio 183 of the Deutsche Forschungsgemeinschaft. 
\end{acknowledgments}

\bibliography{ref}

\clearpage

\onecolumngrid

\def\theequation{S\arabic{equation}}
\def\thefigure{S\arabic{figure}}
\def\thesection{S\arabic{section}}
\setcounter{equation}{0}
\setcounter{figure}{0}
\setcounter{section}{0}

\begin{center}
    {\bf \large Supplementary Materials}    
\end{center}

\section{Considering other disorder terms in the classic percolation argument}

In the main text we have considered the dimerization disorder in the layer construction argument of delocalized states in axion insulator. 
Now we show that considering more complicated disorder terms will not change the conclusion. 
In the decoupled layer limit, changing the chemical potential will lead to the quantum Hall transition in each layer. 
Hence the chemical potential disorder, which does not couple layers, will not localize the states as the quantum Hall transition in each layer is unavoidable. 
The only way to trivialize the Chern layers is to couple even number of layers with opposite Chern numbers.
In the main text, we have considered the dimerization of layers. 
Here we consider an additional disorder that couples four layers.
There are four types of four-layer-couplings: (i) the term coupling the layer at $z=2n$ ($n\in\mathbb{Z}$) to layers at $z=2n-\frac12$, $2n-1$, $2n-\frac32$, (ii) the term coupling the layer at $z=2n-\frac12$ to $z=2n-1$, $2n-\frac32$, $2n-2$, (iii)  the term coupling the layer at $z=2n-1$ to $z=2n-\frac32$, $2n-2$, $2n-\frac52$, (iv)  the term coupling the layer at $z=2n-\frac32$ to $z=2n-2$, $2n-\frac52$, $2n-3$. 
Under the inversions at $z=0$, term (i) and term (ii) transform to each other, and term (iii) and (iv) transform to each other. 
We refer to the regions with terms (i), (ii), (iii), (iv) as phase-I$^\prime$, phase-II$^\prime$, phase-I$^{\prime\prime}$, phase-II$^{\prime\prime}$, respectively. 
Recall that we have defined the regions with dimerizations between layers at $z=n$ and $z=n-\frac12$ ($z=n-\frac12$ and $z=n-1$) as phase-I (phase-II). 
One can find that on the boundary between a phase in \{phase-I,  phase-I$^\prime$,  phase-I$^{\prime\prime}$\} and another phase in  \{phase-II,  phase-II$^\prime$,  phase-II$^{\prime\prime}$\} there must be an odd number of Chern layers.
Since the nearby layers have opposite Chern numbers, the total Chern number on the boundary must be $\pm1$. 
Since  \{phase-I,  phase-I$^\prime$,  phase-I$^{\prime\prime}$\} transform to \{phase-II,  phase-II$^\prime$,  phase-II$^{\prime\prime}$\} under the inversion, we can simply replace the phase-I and phase-II in Fig. 1f by the collections of phases  \{phase-I,  phase-I$^\prime$,  phase-I$^{\prime\prime}$\} and \{phase-II,  phase-II$^\prime$,  phase-II$^{\prime\prime}$\}, respectively.
Then the same argument for the percolating domain walls applies as long as the inversion is recovered on average. 
This argument can be easily to generalized to any disorder terms that couple even number of layers.

\section{Transfer matrix in the \texorpdfstring{$x$-direction}{x-direction}}\label{app:transfer-x}
\begin{figure}[t]
\centering
\includegraphics[width=0.8\linewidth]{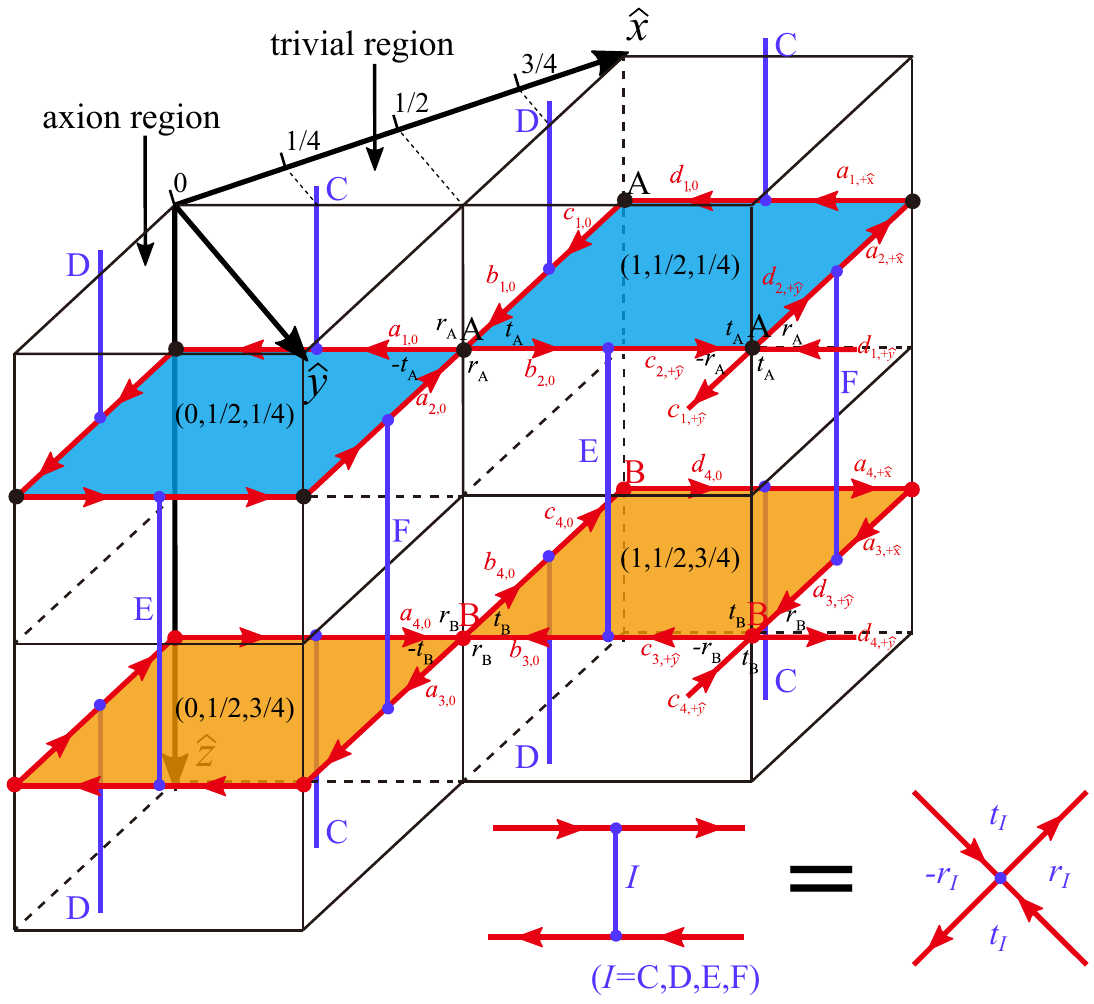}
\caption{Scattering parameters in the quantum network model of disordered axion state.
The unit vectors along the three directions are $\hx,\hy,\hz$, respectively.
Scattering parameters in all the other units are the same.
The chiral modes are labeled as $a_{\alpha j}$, $b_{\alpha j}$, $c_{\alpha j}$, $d_{\alpha j}$, where $\alpha=1,2,3,4$ and $j=(t_x,t_y,t_z)$ ($t_{x,y,z}\in\mathbb{Z}$).
$j=0$ is a shorthand for $j=(000)$.
$\alpha=1,2$ correspond to modes in the blue layer and $\alpha=3,4$ correspond to modes in the orange layer.
The positions $(x,\frac12,\frac14)$ with $x=0,1$ printed on the blue regions are the centers of the corresponding $C=1$ regions, and the positions $(x,\frac12,\frac34)$ with $x=0,1$ printed on the orange regions are the centers of the corresponding $C=-1$ regions.
There are two A nodes in each repeating unit: one scatters $a_{2,j}$ and $b_{1,j}$ to $a_{1,j}$ and $b_{2,j}$, the other scatters  $c_{2,j}$ and $d_{1,j}$ to $c_{1,j}$ and $d_{2,j}$.
There are also two B nodes in each repeating unit: one scatters $a_{4,j}$ and $b_{3,j}$ to $a_{3,j}$ and $b_{4,j}$, the other scatters  $c_{4,j}$ and $d_{3,j}$ to $c_{3,j}$ and $d_{d,j}$.
The node C scatters $a_{1,j+\hat{x}}$ and $d_{4,j-\hat{z}}$ to $a_{1,j+\hat{x}-\hat{z}}$ and $d_{1,j}$.
The node D scatters $b_{4,j-\hat{z}}$ and $c_{1,j}$ to $b_{1,j}$ and $c_{4,j-\hat{z}}$.
The node E scatters $b_{2,j}$ and $c_{3,j+\hat{y}}$ to $b_{3,j}$ and $c_{2,j+\hat{y}}$.
The node F scatters $a_{3,j+\hat{x}}$ and $d_{2,j+\hat{y}}$ to $a_{2,j+\hat{x}}$ and $d_{3,j+\hat{y}}$.
}
\label{fig:transfer3Dx}
\end{figure}

In this section, we construct the transfer matrix in the $x$-direction of the quantum network model shown in Fig. 2 in the main text.
For convenience, we choose the length of the repeating unit as $1$ such that the scattering nodes  are projected to $x=n+\frac14,n+\frac12,n+\frac34,n+1$ ($n\in \mbb{Z}$) (\cref{fig:transfer3Dx}).
We label the modes in the intervals $n+\frac14 < x<n+\frac12$, $n+\frac12 < x<n+\frac34$, $n+\frac34 < x<n+1$, $n+1 < x<n+\frac54$ as $a_{\alpha,j}$, $b_{\alpha,j}$, $c_{\alpha,j}$, $d_{\alpha,j}$, respectively, with $\alpha$ being 1,2 (3,4) for the modes in the blue (orange) layer in each repeating unit and $j=(t_x,t_y,t_z)$ ($t_{x,y,z}\in\mathbb{Z}$) being the lattice index.
The single-slice transfer matrix maps $a_{\alpha,j}$ to $a_{\alpha,j+\hat{x}}$, \ie
\begin{equation}
a_{\alpha, (n+1,t_y,t_z)} = \sum_{\beta t_y' t_z'} T_{\alpha, t_y,t_z ;\  \beta t_y' t_z' }^{(n)} a_{\beta,(n,t_y',t_z')}. \label{eq:T-x}
\end{equation}

Let us first consider the modes $a_{1,j}$, $a_{2,j}$, $b_{1,j}$, $b_{2,j}$ connected to the scattering node A (\cref{fig:transfer3Dx}).
For simplicity, we omit the random phases on the chiral modes for now.
The four modes satisfy
\begin{equation} \label{eq:A-ab-12}
a_{1,j} = -t_A a_{2,j} + r_A b_{1,j},\qquad
b_{2,j} = r_A a_{2,j} + t_{A} b_{1,j},
\end{equation}
where $t_A,r_A$ are the parameters of the $A$ node.
One can then write $b_{1,j}$ and $b_{2,j}$ in terms of $a_{1,j}$ and $a_{2,j}$ as
\begin{equation}
b_{1,j} = \frac1{r_A}a_{1,j} + \frac{t_A}{r_A}a_{2,j},\qquad
b_{2,j} = \frac{t_A}{r_A}a_{1,j} + \frac{1}{r_A}a_{2,j}.
\end{equation}
We can rewrite the above equation as 
$\begin{pmatrix}
b_{1,j} \\ b_{2,j}
\end{pmatrix} = T_A
\begin{pmatrix}
a_{1,j} \\ a_{2,j}
\end{pmatrix}
$ with 
$ T_A=\begin{pmatrix}
\frac1{r_A} & \frac{t_A}{r_A} \\
\frac{t_A}{r_A} & \frac{1}{r_A}
\end{pmatrix} $ 
being the transfer matrix (without random phase) of the node A. 
Now we turn on the (quenched) random phases: We replace $T_A$ by 
$
\begin{pmatrix}
    e^{i\phi^b_{1,j}} & 0 \\
    0 & e^{i\phi^b_{2,j}}
\end{pmatrix}
T_A
$
such that $b_{1,j}$ and $b_{2,j}$ obtain two additional phases
\begin{equation}
b_{1,j} = e^{i\phi^b_{1,j}} \pare{  \frac1{r_A}a_{1,j} + \frac{t_A}{r_A}a_{2,j} },\qquad
b_{2,j} = e^{i\phi^b_{2,j}} \pare{  \frac{t_A}{r_A}a_{1,j} + \frac{1}{r_A}a_{2,j} },
\end{equation}
where $\phi^b_{\alpha j}$ are the random phase factors assigned to $b_{\alpha,j}$.
Similarly, by assigning random phases to the transfer matrices for all the scattering nodes, one can obtain
\begin{equation}
b_{3,j} = e^{i\phi^b_{3,j}} \pare{  \frac1{r_B}a_{3,j} + \frac{t_B}{r_B}a_{4,j} },\qquad
b_{4,j} = e^{i\phi^b_{4,j}} \pare{  \frac{t_B}{r_B}a_{3,j} + \frac{1}{r_B}a_{4,j} },
\end{equation}
\begin{equation}
c_{1,j} = e^{i\phi^c_{1,j}}\pare{\frac1{t_D}b_{1,j} + \frac{r_D}{t_D}b_{4,j-\hz} },\qquad
c_{4,j-\hz} = e^{i\phi^c_{4,j-\hz}}\pare{\frac{r_D}{t_D}b_{1,j} + \frac1{t_D}b_{4,j-\hz}},
\end{equation}
\begin{equation}
c_{2,j+\hy} = e^{i\phi^c_{2,j+\hy}}\pare{\frac1{t_E} b_{2,j} + \frac{r_E}{t_E} b_{3,j}},\qquad
c_{3,j+\hy} = e^{i\phi^c_{3,j+\hy}}\pare{\frac{r_E}{t_E} b_{2,j} + \frac{1}{t_E} b_{3,j}},
\end{equation}
\begin{equation}
d_{1,j} = e^{i\phi^d_{1,j}}\pare{\frac1{t_A} c_{1,j} + \frac{r_A}{t_A} c_{2,j}},\qquad
d_{2,j} = e^{i\phi^d_{2,j}}\pare{\frac{r_A}{t_A} c_{1,j} + \frac{1}{t_A} c_{2,j}},
\end{equation}
\begin{equation}
d_{3,j} = e^{i\phi^d_{3,j}}\pare{\frac1{t_B} c_{3,j} + \frac{r_B}{t_B} c_{4,j}},\qquad
d_{4,j} = e^{i\phi^d_{4,j}}\pare{\frac{r_B}{t_B} c_{3,j} + \frac{1}{t_B} c_{4,j}},
\end{equation}
\begin{equation}
a_{1,j+\hat{x}} = e^{i\phi^a_{1,j+\hx}}\pare{\frac{1}{t_C}d_{1,j} + \frac{r_C}{t_C} d_{4,j-\hz} },\qquad
a_{4,j+\hat{x}-\hz} = e^{i\phi^a_{4,j+\hx-\hz}}\pare{\frac{r_C}{t_C}d_{1,j} + \frac{1}{t_C} d_{4,j-\hz} },
\end{equation}
\begin{equation}
a_{2,j+\hx-\hy} = e^{i\phi^a_{1,j+\hx-\hy}}\pare{\frac{1}{t_F}d_{2,j} + \frac{r_F}{t_F} d_{3,j} },\qquad
a_{3,j+\hx-\hy} = e^{i\phi^a_{3,j+\hx-\hy}}\pare{\frac{r_F}{t_F}d_{2,j} + \frac{1}{t_F} d_{3,j} }.
\end{equation}
Here $\phi^{a,b,c,d}_{\alpha j}$ are the phase factors assigned to $a_{\alpha,j}$ $b_{\alpha,j}$, $c_{\alpha,j}$, $d_{\alpha,j}$, respectively, and $t_{I}$, $r_I$ ($I=A,B,C,D,E,F$) are defined in \cref{fig:transfer3Dx}.
Combining the above equations, we can write $a_{\alpha,j+\hx}$ as linear functions of $a_{\alpha,j}$.
The single-slice transfer matrix $T^{(n)}$ (\cref{eq:T-x}) can then be determined.

\section{Localized limits} \label{app:limits}

In this section, we discuss the localized limits, where electrons are strictly confined in certain regions.

First, we consider the inversion-symmetric trivial insulator limit, where $\eta=\delta=0$ and $\mu=-\frac{\pi}4$.
We have $t_{A,B}=1$ and $r_{A,B}=0$ according to Eqs. (2) and (3) in the main text.
Without inter-layer couplings ($\gamma=0$), we find that the modes form two loops (Fig. 2d in the main text):
\begin{enumerate}
\item $a_{1,j+\hx} \to d_{1,j} \to c_{1,j} \to b_{1,j} \to b_{2,j} \to c_{2,j+\hy} \to d_{2,j+\hy} \to a_{2,j+\hx}$. The last mode connects to the first mode.
\item $a_{3,j+\hx} \to d_{3,j+\hy} \to c_{3,j+\hy} \to b_{3,j} \to b_{4,j} \to c_{4,j} \to d_{4,j} \to a_{4,j+\hx}$. The last mode connects to the first mode.
\end{enumerate}
The first loop flows surrounding the blue Chern insulator, and the second loop flows surrounding the  orange Chern insulator.
The centers of the two loops are $(1,\frac12,\frac14)$ and  $(1,\frac12,\frac34)$, respectively.
With finite inter-layer coupling ($\gamma\neq0,\pi$), the loops centered at $(1,\frac12,\pm \frac14+n)$ ($n\in\mathbb{Z}$) form an infinite long cylinder in the $z$-direction (Fig. 2d in the main text).
This cylinder can be thought as an axion region because it contains Chern layers occupying the inversion centers.
Since the cylinders in different unit cells in the $xy$ directions are disconnected from each other, we can then adiabatically, \ie without creating extended states, shrink them to zero such that the trivial regions fill the whole system.
Thus the localized limit $\mu=-\frac{\pi}4$ corresponds to the trivial insulator phase.

\begin{figure}[]
\centering
\includegraphics[width=0.4\linewidth]{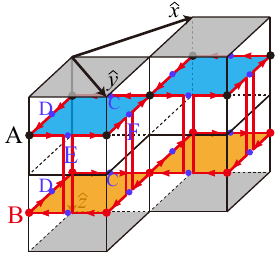}
\caption{ Localization due to inversion symmetry breaking on average, where $t_C=t_D=1$, $t_E=t_F=0$, $t_A=t_B=\frac1{\sqrt2}$. The chiral modes in the slab $0\le z<1$ are reflected back at the C, D nodes in the grey planes ($z=0,1$), hence the network model is decoupled in the $z$-direction.
Each slab has a vanishing Chern number.
}
\label{fig:limits}
\end{figure}

Second, we consider the inversion-symmetric axion insulator limit, where $\eta=\delta=0$ and $\mu=\frac{\pi}4$.
We have $t_{A,B}=0$ and $r_{A,B}=1$ according to Eqs. (2) and (3) in the main text.
Without inter-layer couplings ($\gamma=0$), we find that the modes form two loops in each repeating unit:
\begin{enumerate}
\item $a_{2,j} \to b_{2,j} \to c_{2,j+\hy} \to c_{1,j+\hy} \to b_{1,j+\hy} \to a_{1,j+\hy} \to d_{1,j+\hy-\hx} \to d_{2,j+\hy-\hx}$. The last mode connects to the first mode.
\item $a_{3,j} \to d_{3,j+\hy-\hx} \to d_{4,j+\hy-\hx} \to a_{4,j+\hy} \to b_{4,j+\hy} \to c_{4,j+\hy} \to c_{3,j+\hy} \to b_{3,j}$. The last mode connects to the first mode.
\end{enumerate}
In each repeating unit, the first loop flows surrounding the trivial insulator (empty region) in the layer $z=\frac14$, and the second loop flows surrounding the trivial insulator (empty region) in the layer $z=\frac34$, as shown in Fig. 2e in the main text.
The centers of the two loops are $(\frac12,1,\frac14)$ and  $(\frac12,1,\frac34)$, respectively.
With finite inter-layer coupling ($\gamma\neq0,\pi$), the loops centered at $(\frac12,1,\pm \frac14+n)$ ($n\in\mathbb{Z}$) form an infinite long cylinder in the $z$-direction (Fig. 2e in the main text).
This cylinder can be thought as a trivial insulator region because it does not contain Chern layers.
Since the cylinders in different unit cells in the $xy$ directions are disconnected from each other, we can then adiabatically, \ie without creating extended states, shrink them to zero such that the axion regions fill the whole system.
Thus the localized limit $\mu=\frac{\pi}4$ corresponds to the axion insulator phase.

The case $\mu=\eta=0$ corresponds to the middle of the topological band, where the electronic states are guaranteed to be delocalized if inversion symmetry is respected on average ($\delta=0$), as shown in the phase diagram (Fig. 4a in the main text).
Now we show how symmetry breaking on average ($\delta\neq0$) can lead to localization of these extended states.
We consider $\delta=1$.
We then have $\theta_C=\theta_D=0$, $\theta_E=\theta_F=\frac{\pi}2$ such that $t_C=t_D=1$, $t_E=t_F=0$.
As shown in \cref{fig:limits}, the modes in $z\ge 0 $ cannot tunnel to the modes in $z<0$ since $t_C=t_D=0$.
Thus the modes in $z\ge 1$ are completely decoupled from the modes in $z<0$.
For the same reason, the modes in $z \ge n$ ($n\in \mathbb{Z}$) are completely decoupled from the modes in $z< n$.
Thus the network model is decoupled into a set of 2D slabs: $n-1 \le z < n$.
Each slab has a vanishing Chern number as the blue and orange layers have opposite Chern numbers, and there is no protecting symmetry, hence each slab becomes localized in the presence of disorder.

\section{Numerical methods} \label{app:numeric}

The localization length of the network model can be computed from the transfer matrix with a quasi-1D geometry \cite{mackinnon_one-parameter_1981,chalker_percolation_1988, mackinnon1983scaling}.
We take the $x$-direction as the quasi-1D direction and set the system size as $M\times L\times L $, where $M$ is a large number ($200000$) and $L$ a small number ($\le16$).
We take open boundary condition in the $x$-direction and periodic boundary condition in the $y,z$-directions.
The amplitudes of the chiral modes in the $x=n+1$ slice is determined by the amplitudes in the $x=n$ slice  through the single-slice transfer matrix \cref{eq:T-x}.
The amplitudes in the $x=M$ slice are then $a^{(x=M)}=T_M a^{(x=1)}$, where $T_M=T^{(M)}T^{(M-1)}\cdots T^{(1)}$ for a given finite $L$.
Here each $T^{(i)}$ is a transfer matrix for a single slice with random phases, and the random phases in different slices are independent.
The theorem by Oseledec \cite{Ose1968} guarantees that the eigenvalues of the following matrix converge to finite values
\begin{equation}
\Gamma = \lim_{M\to \infty} (T_{M}^\dagger T_M)^{\frac{1}{2M}}.\label{eq:Gamma}
\end{equation}
As proven in Ref. \cite{Ose1968}, the eigenvalues of $\Gamma$ come in pairs of $e^{\pm \gamma_i}$ ($\gamma_i>0$).
As explained below in this paragraph, physically one can understand $e^{\gamma_i}$ and $e^{-\gamma_i}$ as the growing and decaying channels of the transfer matrix.
Given finite $a_{\alpha,j}$ ($\sim 1$) in the $x=0$ slice, then the amplitudes in the $x=M$ slice are of order $\Gamma^M a_{\alpha,j} \sim e^{\pm \gamma_i M}$, wherein the positive and negative signs represent the growing and decaying channels, respectively.
Thus, the inverse of the minimal value of $\gamma_i$, \ie $\min\{\gamma_i\}^{-1}$, gives the localization length of the quasi-1D system.

The eigenvalues of $\Gamma$ cannot be computed directly from the definition \cref{eq:Gamma} because $T^{\dagger}_M T_M$ is a divergent matrix.
Now we discuss the method to compute the eigenvalues. 
First we write $T_M$ as a product of $M_1$ super transfer matrices
\begin{equation}
T_{M} = T^{(M_1) \pr} T^{(M_1-1) \pr} \cdots T^{(1) \pr},
\end{equation}
where a single super transfer matrix is a product of $M_0$ transfer matrices
\begin{equation}
T^{(k)\pr} = T^{(kM_0)} T^{(kM_0-1)} \cdots T^{((k-1)M_0+1)}.
\end{equation}
The total length of the quasi-1D system is $M=M_1M_0$.
We choose $M_0$ as such a small number that the super transfer matrices do not numerically diverge.
In practice, we choose $M_0=5$.
However, we have numerical difficulty to evaluate the products of the super transfer matrices because the largest eigenvalue of the product grows exponentially with $M$.
To overcome this difficulty, we only store the logarithms of the amplitudes in each step.
We apply a QR-decomposition of $T^{(1)\pr}$ as
$T^{(1)\pr} = Q_1 R_1$, with $Q_1$ being a unitary matrix and $R_1$ a upper trigonal matrix.
We rewrite $T^M$ as
\begin{equation}
    T_M = T^{(M_1) \pr} T^{(M_1-1) \pr} \cdots T^{(2) \pr} Q_1 R_1.
\end{equation}
Then we apply the QR-decomposition $T^{(2)\pr} Q_1 = Q_2 R_2$ and obtain
\begin{equation}
    T_{M} = T^{(M_1) \pr} T^{(M_1-1) \pr} \cdots T^{(3) \pr} Q_2 R_2 R_1.
\end{equation}
Continuing this procedure, we obtain
\begin{equation}
T_M = Q_{M_1} R_{M_1} \cdots R_2 R_1.
\end{equation}
Ref.~\cite{mackinnon1983scaling} showed that the eigenvalues of $\Gamma^M$ have the same magnitude orders as the diagonal elements of $R_{M_1} \cdots R_2 R_1$.
Since $R_{M_1\cdots 1}$ are upper trigonal, the diagonal elements of $R_{M_1} \cdots R_2 R_1$ are given as the products of those of the $R_k$ ($k=1\cdots M_1$) matrices.
Therefore, we can approximate the eigenvalues as
\begin{equation}
\gamma_i = \frac{1}{M} \sum_{k=1}^{M_1} \ln( [R_k]_{ii} ). \label{eq:gamma-def}
\end{equation}
And the uncertainty of $\gamma_i$ is given by
\begin{equation}
\delta \gamma_i = \sqrt{\frac1{M} \sum_{k=1}^{M_1} \abs{\ln([R_k]_{ii})}^2 -  \pare{\frac{1}{M} \sum_{k=1}^{M_1} \ln( [R_k]_{ii} )}^2 } \ .\label{eq:dgamm}
\end{equation}
In practice, one only need to store and sum the logarithms of the diagonal elements of the $R$ matrices.
In the following, we will denote the smallest $\gamma_i$ and its uncertainty as $\gamma_{\rm min}$ and $\delta \gamma_{\rm \min}$ for simplicity, respectively.

In the above we have introduced the method to compute $\gamma_{\rm min}$ for a quasi-1D geometry with the size $M\times L\times L$.
We call such a quasi-1D sample as a single disorder configuration, where the transfer matrices for different slices have independent random phases.
To further reduce the uncertainty $\delta\gamma_{\rm min}$, we have generated $N_D=28$ different disorder configurations for each $L$ and $M$, and take the average $\gamma_{\rm min}$ over the 28 configurations. 
We denote the $\gamma_{\rm min}$'s and their uncertainties from different disorder configurations as $\gamma_{\rm min}^{(n)}$ and $\delta \gamma_{\rm min}^{(n)}$, respectively, with $n=1\cdots N_D$ labeling the disorder configurations.
The averaged $\gamma_{\rm min}$ and its uncertainty are given by
\begin{equation} \label{eq:bargamma}
\bar{\gamma}_{\rm min} = \frac{1}{N_D} \sum_{n=1}^{N_D} \gamma_{\rm min}^{(n)}\ ,\qquad
\delta \bar{\gamma}_{\rm min} = \sqrt{\frac{1}{N_D} \sum_{n=1}^{N_D} \big(\delta \gamma_{\rm min}^{(n)}\big)^2 } \ , 
\end{equation}
respectively.
The localization length of the quasi-1D system is $\lambda=1/\bar{\gamma}_{\rm min}$.
The normalized localization length used in the main text is defined as $\Lambda=\lambda/L$. 
The 3D localization length is given by $\xi=\lim_{L\to\infty}\lambda$.

We apply the least square method to fit the parameters in Eq. 4 in the main text.
There are 11 parameters to be fitted: $\mu_c$, $\nu$, $y$, $g^{(i)}_m$ ($i=0,1$, $m=0,1,2,3$).
For convenience, we relabel these 11 parameters as
\begin{equation}
\begin{split}
& \zeta_{1} \equiv \mu_c,\quad
\zeta_2 \equiv \nu,\quad
\zeta_3 \equiv y,\quad
\zeta_4 \equiv g^{(0)}_0,\quad
\zeta_5 \equiv g^{(0)}_1,\quad
\zeta_6 \equiv g^{(0)}_2,\quad
\zeta_7 \equiv g^{(0)}_3,\\
& \zeta_8 \equiv g^{(1)}_0,\quad
\zeta_9 \equiv g^{(1)}_1,\quad
\zeta_{10} \equiv g^{(1)}_2,\quad
\zeta_{11} \equiv g^{(1)}_3.
\end{split} \label{eq:zeta-def}
\end{equation}
We use $\Lambda(\mu,L,\{\zeta_i\})$ to represent the $\Lambda$ calculated as a function of $\mu,L$ using Eq. 4 in the main text, \ie
\begin{equation}
\Lambda(\mu,L,\{\zeta_i\}) = \sum_{m=0}^3 g_m^{(0)} ((\mu-\mu_c) L^{\frac1{\nu}})^m + L^{y} \sum_{m=0}^3 g_m^{(1)} ((\mu-\mu_c) L^{\frac1{\nu}})^m.
\end{equation}
We use $\{\mu_n, L_n, \Lambda_n \}$ ($n=1\cdots N$) to represent numerical results where $\Lambda_n$ is given by the transfer matrix method with the chemical potential $\mu=\mu_n$ and transverse size $L=L_n$.
$N$ is the size the data.
We apply the conjugate gradient method to minimize the error
\begin{equation}
S(\{\zeta_i\}) = \frac1{N} \sum_{n} ( \Lambda_n - \Lambda(\mu_n,L_n,\{\zeta_i\}))^2.
\end{equation}
The conjugate gradient algorithm needs the first order and second order derivatives of $S$.
We compute the first order derivative as
\begin{equation}
\frac{\partial S}{\partial \zeta_i} = - \frac2{N} \sum_n ( \Lambda_n - \Lambda(\mu_n,L_n,\{\zeta_i\}))  \frac{\partial \Lambda(\mu_n,L_n,\{\zeta_i\})}{\partial \zeta_i}.
\end{equation}
The second order derivative is
\begin{equation}
\frac{\partial^2 S}{\partial \zeta_i \partial \zeta_j} = -\frac2{N} \sum_n ( \Lambda_n - \Lambda(\mu_n,L_n,\{\zeta_i\}))  \frac{\partial^2  \Lambda(\mu_n,L_n,\{\zeta_i\})}{\partial \zeta_i \partial \zeta_j} + \frac2{N} \sum_n \frac{\partial \Lambda(\mu_n,L_n,\{\zeta_i\})}{\partial \zeta_i} \frac{\partial \Lambda(\mu_n,L_n,\{\zeta_i\})}{\partial \zeta_j}.
\end{equation}
Since $\Lambda(\mu_n,L_n,\{\zeta_i\}$ is fitted to match $\Lambda_n$, we can roughly regard $\Lambda(\mu_n,L_n,\{\zeta_i\})$ as the average of $\Lambda_n$ and hence we can regard $\Lambda_n$ as a random number centered at $\Lambda(\mu_n,L_n,\{\zeta_i\})$, whose uncertainty origins from the numerical uncertainty of $\gamma_i$. 
When $S$ is small, the term $( \Lambda_n - \Lambda(\mu_n,L_n,\{\zeta_i\}))$ can be thought as a small random number centered at zero.
Summing over sufficient many random numbers cancels the first term.
Thus the second order derivative can be well approximated as \cite{press_numerical_1996}
\begin{equation}
\frac{\partial^2 S}{\partial \zeta_i \partial \zeta_j} \approx \frac2{N} \sum_n \frac{\partial \Lambda(\mu_n,L_n,\{\zeta_i\})}{\partial \zeta_i} \frac{\partial \Lambda(\mu_n,L_n,\{\zeta_i\})}{\partial \zeta_j}.
\end{equation}
In our calculation, there are about 50 data points around the transition point to be fitted.
For example, in Fig. 3 in the main text, we have used 49 data points in the fitting, as marked by the dashed square in Fig. 3a in the main text.
For all the fittings, the error is minimized to be smaller than $10^{-4}$.

We have re-sampled the numerical data $\{\mu_n,L_n,\Lambda_n\}$ to obtain the error-bars of the fitted parameters.
For each $\mu_n$ and $L_n$, we can generate a new normalized localization length $\td{\Lambda}_n=1/(\td{\gamma}_{\rm min}L_n)$, where $\td{\gamma}_{\rm min}$ is a gaussian random number with the uncertainty and expectation given in \cref{eq:bargamma}.
Then we apply the fitting scheme introduced above for the newly generated data set $\{\mu_n, L_n, \td{\Lambda}_n\}$ to obtain a new set of the fitted parameters (\cref{eq:zeta-def}).
Repeating the re-sampling and fitting for many times (200 in our work), we can count the uncertainties in the fitted parameters.

\section{Gap-closing transition from trivial insulator to axion insulator}\label{app:gap}
For a band structure with inversion symmetry in the absence of disorder, one can define the indices \cite{Fu2007Inversion,Huse2011Inversion,Ashvin2012Axion,watanabe_structure_2018,MTQC}
\begin{equation}
z_{2,i} = \sum_{\substack{K\\K_i=\pi}} n_{K}^- \mod 2,\qquad i=x,y,z\label{eq:z2}
\end{equation}
\begin{equation}
z_4 = \sum_{K} n_K^- \mod 4 = \sum_{K} \frac{n_K^- - n_K^+}2\mod 4, \label{eq:z4}
\end{equation}
where $K$ indexes all the inversion-invariant momenta, and $n_K^-$ ($n_K^+$) is the number of odd (even) occupied Bloch states at $K$.
The values $z_4=1,3$ correspond to WSM with an odd number of pairs of Weyl nodes.
Provided that the band structure is {\it gapped}, the values $z_4=0,2$ correspond to trivial insulator, 3D QAH insulator, or axion insulator: if $z_{2,i}=1$, then the band is a 3D QAH state with odd weak Chern numbers in the $i$-direction; if $z_{2,i=x,y,z}=z_4=0$, then the band is a trivial insulator or a 3D QAH state with even weak Chern numbers; if $z_{2,i=x,y,z}=0$ and $z_4=2$, then the band is an axion insulator or a 3D QAH state with even weak Chern numbers.
(If the $k_i=0$ and $k_i=\pi$ planes have different Chern numbers, which can be implied by $z_4=1,3$ when the difference is odd, then the band structure has equal number of un-avoidable Weyl points between the $k_i=0$ and $k_i=\pi$ planes to the difference of Chern numbers.)
Now we look at the low energy band structure around $\Gamma$ shown in Fig. 4b in the main text.
We assume that the weak Chern numbers are zero and that other inversion-invariant momenta do not contribute to $z_4$.
For $M>0$, the lower two bands have $z_4=0$ and hence form a trivial insulator.
For $-\Delta<M<0$, the lower two bands have $z_4=1$ and a single pair of Weyl nodes between the highest occupied band the lowest empty band is generated \cite{Ashvin2012Axion}.
For $M<-\Delta$, the lower two bands have $z_4=2$.
At $M=-\Delta$ another pair of Weyl nodes are created.
We assume at some critical $M_c$ ($M_c<-\Delta$) the two pairs of Weyl nodes annihilate each other.
Then for $M<M_c$ the lower two bands become an axion insulator.

If additional symmetries are present, the two occupied (empty) states at $\Gamma$ may become degenerate.
For example, the additional symmetries can be $m_x$ ($x\to -x$) and $m_y$ ($y\to -y$) with non-negligible spin-orbit coupling.
The anti-commuting relation $\{m_x,m_y\}=0$ guarantees that the states at $\Gamma$ are doubly degenerate, and the two states in each doublet have the same parity.
Then there must be $\Delta=0$ and the $z_4$ index will change from 0 to 2 for $M$ changing from positive to negative.
The intermediate WSM phase is not un-avoidable in this phase transition process.
Whether a WSM phase appears in the phase transition depends on the details of the band structure.
When the WSM does appear, it must have an even number of pairs of Weyl points since $z_4=0,2$. 

\section{Mapping the quantum network model to an effective Hamiltonian}
\label{app:mapping}

We map the quantum network model to a Floquet system: at each scattering node the incoming modes evolve to the outgoing modes after a time $\Delta t$.
We follow the method introduced in Ref. \cite{Ho1996network}.
The disordered phases $\phi_{\alpha,j}^{a,b,c,d}$ will be mapped to the disordered vector potential of a disordered magnetic field in the effective Hamiltonian of the Floquet system \cite{wang_universal_2014}, because the phases $\phi_{\alpha,j}^{a,b,c,d}$ introduce disordered fluxes in each loop enclosed by the chiral modes. 
In the following, we only consider the clean limit ($\phi_{\alpha,j}^{a,b,c,d}=0$) for simplicity, which corresponds to vanishing disordered magnetic field. 
The evolution operator from modes $a_{i,\alpha}$ to $a_{j,\beta}$ ($\alpha,\beta=1,2,3,4$) is
\begin{equation}
U_{aj\beta,ai\alpha} = \begin{pmatrix}
0 & -t_A \delta_{ji} & 0 & 0\\
0 &  0   & r_F \delta_{ji} & 0\\
0 & 0 & 0 & -t_B \delta_{ji} \\
r_C [T_z^{-1}]_{j,i} & 0 & 0 & 0
\end{pmatrix}_{\beta,\alpha} \ ,
\end{equation}
where the columns and the rows correspond to the initial and final states, respectively.
$[T_z]_{j,i}=\delta_{j,i+\hz}$ is the translation operator in the $z$-direction.
The matrix elements can be read from the scattering nodes in \cref{fig:transfer3Dx}.
For examples, since $a_{1,j}=-t_A a_{2,j} + r_A b_{1,j}$ (\cref{eq:A-ab-12}), we have $U_{aj1,aj2} =-t_A \delta_{ji}$; since $a_{2,j} = r_F a_{3,j} + t_F d_{2,j+\hat{y}-\hat{x}}$, as shown in \cref{fig:transfer3Dx}, we have $U_{aj2,aj3} = r_F \delta_{ji}$. 

The total evolution operator has sixteen blocks: $U_{\mu,\nu}$ ($\mu,\nu=a,b,c,d$).
We will derive them in the next paragraph.
Since all the scattering nodes satisfy $t_I^2+r_I^2=1$ ($I=A\cdots F$), the total current of the chiral states, \ie $\sum_{\alpha=1}^4 \sum_{j} |a_{\alpha,j}|^2 + |b_{\alpha,j}|^2 + |c_{\alpha,j}|^2 + |d_{\alpha,j}|^2$, is preserved in the evolution. 
In other words, $U$ preserves the norm of the magnitudes and hence is unitary.

We apply the Fourier transformation
\begin{equation}
a_{\kk,\alpha} = \frac{1}{\sqrt{N}} \sum_{j} e^{i\kk\cdot(j+\tau_\alpha)} a_{j,\alpha},\qquad
\tau_{1,2}=(0,\frac12,\frac14),\; \tau_{3,4}=(0,\frac12,\frac34),
\end{equation}
where we assume that chiral modes locate at the center of the Chern insulator region that they flow around (\cref{fig:transfer3Dx}).
Then we obtain the evolution operator on the Bloch wavefunction $\{a_{\kk,\alpha}\}$
\begin{equation}
U_{aa}(\kk)=\begin{pmatrix}
0 & -t_A & 0 & 0\\
0 &  0   & r_F e^{-i\frac{k_z}2} & 0\\
0 & 0 & 0 & -t_B \\
r_C e^{-i\frac{k_z}2} & 0 & 0 & 0
\end{pmatrix}.
\end{equation}
One can similarly write down other evolution operator blocks $U_{\mu,\nu}$ for $\mu,\nu=a,b,c,d$ in the clean limit.
Here we only give the evolution operators on the Bloch bases.
The Bloch bases for the $b,c,d$ modes are defined as
\begin{equation}
b_{\kk,\alpha} = \frac{1}{\sqrt{N}} \sum_{j} e^{i\kk\cdot(j+\tau_\alpha)} b_{j,\alpha},\qquad
\tau_{1,2}=(1,\frac12,\frac14),\; \tau_{3,4}=(1,\frac12,\frac34),
\end{equation}
\begin{equation}
c_{\kk,\alpha} = \frac{1}{\sqrt{N}} \sum_{j} e^{i\kk\cdot(j+\tau_\alpha)} c_{j,\alpha},\qquad
\tau_{1}=(1,\frac12,\frac14),\; \tau_{2}=(1,-\frac12,\frac14),\;
\tau_{3}=(1,-\frac12,\frac34),\; \tau_{4}=(1,\frac12,\frac34),
\end{equation}
\begin{equation}
d_{\kk,\alpha} = \frac{1}{\sqrt{N}} \sum_{j} e^{i\kk\cdot(j+\tau_\alpha)} d_{j,\alpha},\qquad
\tau_{1}=(1,\frac12,\frac14),\; \tau_{2}=(1,-\frac12,\frac14),\;
\tau_{3}=(1,-\frac12,\frac34),\; \tau_{4}=(1,\frac12,\frac34),
\end{equation}
respectively.
Then the evolution operator blocks are
{\scriptsize
\begin{equation}
U_{ba}(\kk) = \left(
\begin{array}{cccc}
 0 & 0 & 0 & 0 \\
 0 & e^{i k_x} r_A & 0 & 0 \\
 0 & 0 & 0 & 0 \\
 0 & 0 & 0 & e^{i k_x} r_B \\
\end{array}
\right),\quad
U_{ca}(\kk)=0,\quad
U_{da}(\kk)=\left(
\begin{array}{cccc}
 {t_C} & 0 & 0 & 0 \\
 0 & 0 & 0 & 0 \\
 0 & 0 & {t_F} & 0 \\
 0 & 0 & 0 & 0 \\
\end{array}
\right),
\end{equation}
\begin{equation}
U_{ab}(\kk) = \left(
\begin{array}{cccc}
 e^{-i {k_x}} {r_A} & 0 & 0 & 0 \\
 0 & 0 & 0 & 0 \\
 0 & 0 & e^{-i {k_x}} {r_B} & 0 \\
 0 & 0 & 0 & 0 \\
\end{array}
\right),\quad
U_{bb}(\kk) = \left(
\begin{array}{cccc}
 0 & 0 & 0 & -e^{\frac{i {k_z}}{2}} {r_D} \\
 {t_A} & 0 & 0 & 0 \\
 0 & -e^{\frac{i {k_z}}{2}} {r_E} & 0 & 0 \\
 0 & 0 & {t_B} & 0 \\
\end{array}
\right),\quad
U_{cb}(\kk)=\left(
\begin{array}{cccc}
 0 & 0 & 0 & 0 \\
 0 & {t_E} & 0 & 0 \\
 0 & 0 & 0 & 0 \\
 0 & 0 & 0 & {t_D} \\
\end{array}
\right),\quad
U_{db}(\kk)=0,
\end{equation}
\begin{equation}
U_{ac}(\kk)=0,\quad
U_{bc}(\kk)=\left(
\begin{array}{cccc}
 {t_D} & 0 & 0 & 0 \\
 0 & 0 & 0 & 0 \\
 0 & 0 & {t_E} & 0 \\
 0 & 0 & 0 & 0 \\
\end{array}
\right),\quad
U_{cc}(\kk)=\left(
\begin{array}{cccc}
 0 & -e^{i {k_y}} {r_A} & 0 & 0 \\
 0 & 0 & e^{-\frac{i {k_z}}{2}} {r_E} & 0 \\
 0 & 0 & 0 & -e^{-i {k_y}} {r_B} \\
 e^{-\frac{i {k_z}}{2}} {r_D} & 0 & 0 & 0 \\
\end{array}
\right),\quad
U_{dc}(\kk)=\left(
\begin{array}{cccc}
 0 & 0 & 0 & 0 \\
 0 & {t_A} & 0 & 0 \\
 0 & 0 & 0 & 0 \\
 0 & 0 & 0 & {t_B} \\
\end{array}
\right),
\end{equation}
\begin{equation}
U_{ad}(\kk) = \left(
\begin{array}{cccc}
 0 & 0 & 0 & 0 \\
 0 & {t_F} & 0 & 0 \\
 0 & 0 & 0 & 0 \\
 0 & 0 & 0 & {t_C} \\
\end{array}
\right),\quad
U_{bd}(\kk)=0,\quad
U_{cd}(\kk)=\left(
\begin{array}{cccc}
 {t_A} & 0 & 0 & 0 \\
 0 & 0 & 0 & 0 \\
 0 & 0 & {t_B} & 0 \\
 0 & 0 & 0 & 0 \\
\end{array}
\right),\quad
U_{dd}(\kk)=\left(
\begin{array}{cccc}
 0 & 0 & 0 & -e^{\frac{i {k_z}}{2}} {r_C} \\
 e^{-i {k_y}} {r_A} & 0 & 0 & 0 \\
 0 & -e^{\frac{i {k_z}}{2}} {r_F} & 0 & 0 \\
 0 & 0 & e^{i {k_y}} {r_B} & 0 \\
\end{array}
\right).
\end{equation}
}%
The complete 16-by-16 evolution operator is given by
\begin{equation}
U(\kk)=\begin{pmatrix}
U_{aa}(\kk) & U_{ab}(\kk) & U_{ac}(\kk) & U_{ad}(\kk)\\
U_{ba}(\kk) & U_{bb}(\kk) & U_{bc}(\kk) & U_{bd}(\kk)\\
U_{ca}(\kk) & U_{cb}(\kk) & U_{cc}(\kk) & U_{cd}(\kk)\\
U_{da}(\kk) & U_{db}(\kk) & U_{dc}(\kk) & U_{dd}(\kk)
\end{pmatrix}.
\end{equation}
The inversion operator is
\begin{equation}
\mathcal{I} = \begin{pmatrix}
0 & \tau_0 \otimes \sigma_y & 0 & 0\\
\tau_0 \otimes \sigma_y & 0 & 0 & 0\\
0 & 0 & 0 & \tau_0 \otimes \sigma_y\\
0 & 0 & \tau_0 \otimes \sigma_y  & 0
\end{pmatrix},
\end{equation}
where $\sigma_y$ is the second Pauli-matrix and $\tau_0$ is two-by-two identity matrix.
One can verify that
\begin{equation}
\mathcal{I} U(\kk) \mathcal{I}^{-1} = U(-\kk).
\end{equation}

We can divide the sixteen modes into four groups: $\{a_{\kk,1}, b_{\kk,2}, c_{\kk,3}, d_{\kk,4}\}$, $\{a_{\kk,4}, b_{\kk,3}, c_{\kk,2}, d_{\kk,1}\}$, $\{a_{\kk,3}, b_{\kk,4}, c_{\kk,1}, d_{\kk,2}\}$, $\{a_{\kk,2}, b_{\kk,1}, c_{\kk,4}, d_{\kk,3}\}$.
Then, following  the scattering process in \cref{fig:transfer3Dx}, we find that the modes in the four groups evolve to each other in turn at each step:
\begin{equation}
\begin{array}{ccc}
\{a_{\kk,1}, b_{\kk,2}, c_{\kk,3}, d_{\kk,4}\} & \to & \{a_{\kk,4}, b_{\kk,3}, c_{\kk,2}, d_{\kk,1}\} \\
\uparrow & & \downarrow \\
\{a_{\kk,2}, b_{\kk,1}, c_{\kk,4}, d_{\kk,3}\} & \leftarrow &  \{a_{\kk,3}, b_{\kk,4}, c_{\kk,1}, d_{\kk,2}\}
\end{array} \ .
\end{equation} 
Therefore, the evolution operator for time $4\Delta t$, $U^{4}(\kk)$, must be block-diagonal in the four groups. 
The long-time behavior in the four blocks must be the same because they evolve to each other in a short time $\Delta t$.
Therefore, we only need to look at one block for the low energy physics.
Here we take the block spanned by  $\{a_{\kk,1}, b_{\kk,2}, c_{\kk,3}, d_{\kk,4}\}$ and denote the corresponding evolution operator for time $4\Delta t$ as $\td{U}(\kk)$.
Its matrix elements are
\begin{equation}
\td{U}_{11}(\kk) =  {r_A}  { t_A}  { t_C} \left (e^{-i  {k_x}}  { t_D}-e^{-i  {k_y}}  { t_F}\right)+e^{-i  {k_z}}  {r_C}  {r_F}  { t_A}  { t_B}- {r_A}  {r_B}
{r_C}  {r_D},
\end{equation}
\begin{equation}
\td{U}_{21}(\kk) = {r_A}^2 {t_C} {t_F} e^{i ({k_x}-{k_y})}-{r_A} {r_C} {r_F} {t_B} e^{i ({k_x}-{k_z})}-e^{i {k_x}} {r_B} {r_C} {r_D}  {t_A}+{t_A}^2 {t_C} {t_D},
\end{equation}
\begin{equation}
\td{U}_{31}(\kk) = -e^{-\frac{1}{2} i (2 {k_y}+{k_z})} \left(e^{i {k_x}} {r_B}^2 {r_C} {t_D}+{t_B} \left(e^{i {k_y}} {r_C} {t_B} {t_F}+e^{i
{k_z}} {r_A} {r_F} {t_C}\right)+{r_B} {r_D} {t_A} {t_C}\right),
\end{equation}
\begin{equation}
\td{U}_{41}(\kk) = e^{-\frac{i {k_z}}{2}} \left(e^{i {k_x}} {r_B} {r_C} {t_B} {t_D}-e^{i {k_y}} {r_B} {r_C} {t_B} {t_F}-e^{i {k_z}} {r_A}
{r_B} {r_F} {t_C}+{r_D} {t_A} {t_B} {t_C}\right),
\end{equation}
\begin{equation}
\td{U}_{12}(\kk) = e^{-i {k_x}} \left(-e^{i {k_x}} {t_A}^2 {t_E} {t_F}-e^{i {k_y}} {r_A}^2 {t_D} {t_E}+e^{i {k_z}} {r_A} {r_D} {r_E}
{t_B}+{r_B} {r_E} {r_F} {t_A}\right),
\end{equation}
\begin{equation}
\td{U}_{22}(\kk) = e^{i {k_z}} {r_D} {r_E} {t_A} {t_B}-{r_A} \left({r_B} {r_E} {r_F}+{t_A} {t_E} \left(e^{i {k_y}} {t_D}-e^{i {k_x}}
{t_F}\right)\right),
\end{equation}
\begin{equation}
\td{U}_{32}(\kk) = e^{-\frac{i {k_z}}{2}} \left({t_E} \left({r_A} {r_B} {r_D}-e^{i {k_z}} {r_F} {t_A} {t_B}\right)-e^{i {k_z}} {r_B} {r_E}
{t_B} \left(e^{-i {k_x}} {t_F}-e^{-i {k_y}} {t_D}\right)\right),
\end{equation}
{\small\begin{equation}
\td{U}_{42}(\kk) = e^{-\frac{1}{2} i (2 {k_x}+{k_z})} \left(-{r_B} {r_F} {t_A} {t_E} e^{i ({k_x}+{k_y}+{k_z})}-{r_A} {r_D} {t_B} {t_E} e^{i
({k_x}+{k_y})}-{r_E} {t_B}^2 {t_D} e^{i ({k_x}+{k_z})}+{r_B}^2 {r_E} {t_F} \left(-e^{i ({k_y}+{k_z})}\right)\right),
\end{equation}}%
\begin{equation}
\td{U}_{13}(\kk) = -e^{-\frac{1}{2} i (2 {k_x}+{k_z})} \left(e^{i {k_x}} {r_E} {t_A}^2 {t_F}+e^{i {k_y}} {r_A}^2 {r_E} {t_D}+e^{i {k_z}} {r_A}
{r_D} {t_B} {t_E}+{r_B} {r_F} {t_A} {t_E}\right),
\end{equation}
\begin{equation}
\td{U}_{23}(\kk) = e^{-\frac{i {k_z}}{2}} \left(e^{i {k_x}} {r_A} {r_E} {t_A} {t_F}-e^{i {k_y}} {r_A} {r_E} {t_A} {t_D}-e^{i {k_z}} {r_D}
{t_A} {t_B} {t_E}+{r_A} {r_B} {r_F} {t_E}\right),
\end{equation}
\begin{equation}
\td{U}_{33}(\kk) = e^{-i {k_x}} {r_B} {t_B} {t_E} {t_F}-e^{-i {k_y}} {r_B} {t_B} {t_D} {t_E}+e^{-i {k_z}} {r_A} {r_B} {r_D}
{r_E}-{r_E} {r_F} {t_A} {t_B},
\end{equation}
\begin{equation}
\td{U}_{43}(\kk) = {r_B}^2 {t_E} {t_F} e^{-i ({k_x}-{k_y})}-{r_A} {r_D} {r_E} {t_B} e^{i ({k_y}-{k_z})}-e^{i {k_y}} {r_B} {r_E} {r_F}
{t_A}+{t_B}^2 {t_D} {t_E},
\end{equation}
\begin{equation}
\td{U}_{14}(\kk) = e^{-\frac{i {k_z}}{2}} \left(-e^{i {k_z}} {r_A} {r_C} {t_A} \left(e^{-i {k_x}} {t_D}-e^{-i {k_y}} {t_F}\right)-e^{i {k_z}} {r_A}
{r_B} {r_D} {t_C}+{r_F} {t_A} {t_B} {t_C}\right),
\end{equation}
{\small \begin{equation}
\td{U}_{24}(\kk) = e^{-\frac{1}{2} i (2 {k_y}+{k_z})} \left(-{r_B} {r_D} {t_A} {t_C} e^{i ({k_x}+{k_y}+{k_z})}-{r_A} {r_F} {t_B} {t_C} e^{i
({k_x}+{k_y})}+{r_A}^2 {r_C} {t_F} \left(-e^{i ({k_x}+{k_z})}\right)-{r_C} {t_A}^2 {t_D} e^{i ({k_y}+{k_z})}\right),
\end{equation}}%
\begin{equation}
\td{U}_{34}(\kk) = e^{-i {k_y}} \left(-e^{i {k_x}} {r_B}^2 {t_C} {t_D}+{t_B} \left(e^{i {k_z}} {r_A} {r_C} {r_F}-e^{i {k_y}} {t_B} {t_C}
{t_F}\right)+{r_B} {r_C} {r_D} {t_A}\right),
\end{equation}
\begin{equation}
\td{U}_{44}(\kk) = e^{i {k_x}} {r_B} {t_B} {t_C} {t_D}-e^{i {k_y}} {r_B} {t_B} {t_C} {t_F}+e^{i {k_z}} {r_A} {r_B} {r_C}
{r_F}-{r_C} {r_D} {t_A} {t_B}.
\end{equation}
The projected inversion operator on the bases $\{a_{\kk,1}, b_{\kk,2}, c_{\kk,3}, d_{\kk,4}\}$ is
\begin{equation}
\td{\mathcal{I}} = \tau_0 \otimes \sigma_y. \label{eq:INV4}
\end{equation}

\begin{figure}[t]
\centering
\includegraphics[width=0.9\linewidth]{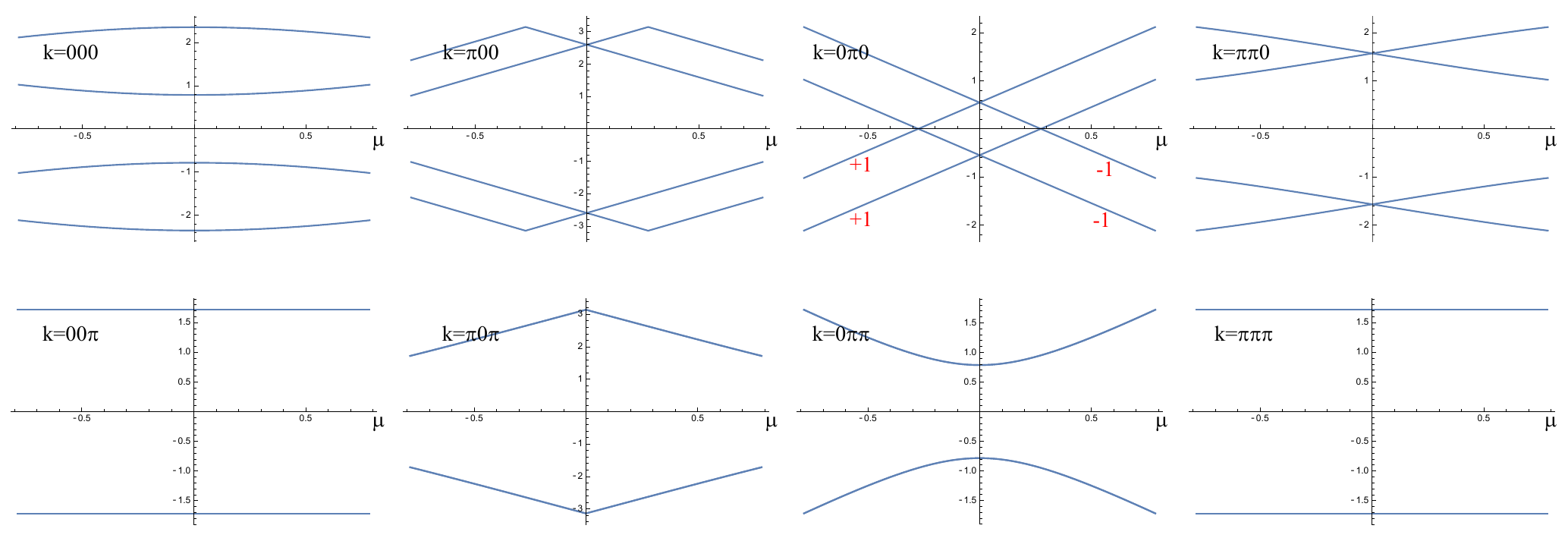}
\caption{The quasi-energies of the effective Hamiltonian $\mathcal{H}(\kk)$ at the eight inversion-invariant momenta as functions of $\mu$ for $-\frac{\pi}4<\mu<\frac{\pi}4$. The other parameters are set to $\eta=\delta=0$, $\gamma=\frac{\pi}8$.
The gap at $\kk=(0\pi0)$ closes at $\mu\approx \pm 0.27$, where the red numbers represent the inversion eigenvalues.}
\label{fig:TBGap-M}
\end{figure}

\begin{figure}[h]
\centering
\includegraphics[width=0.9\linewidth]{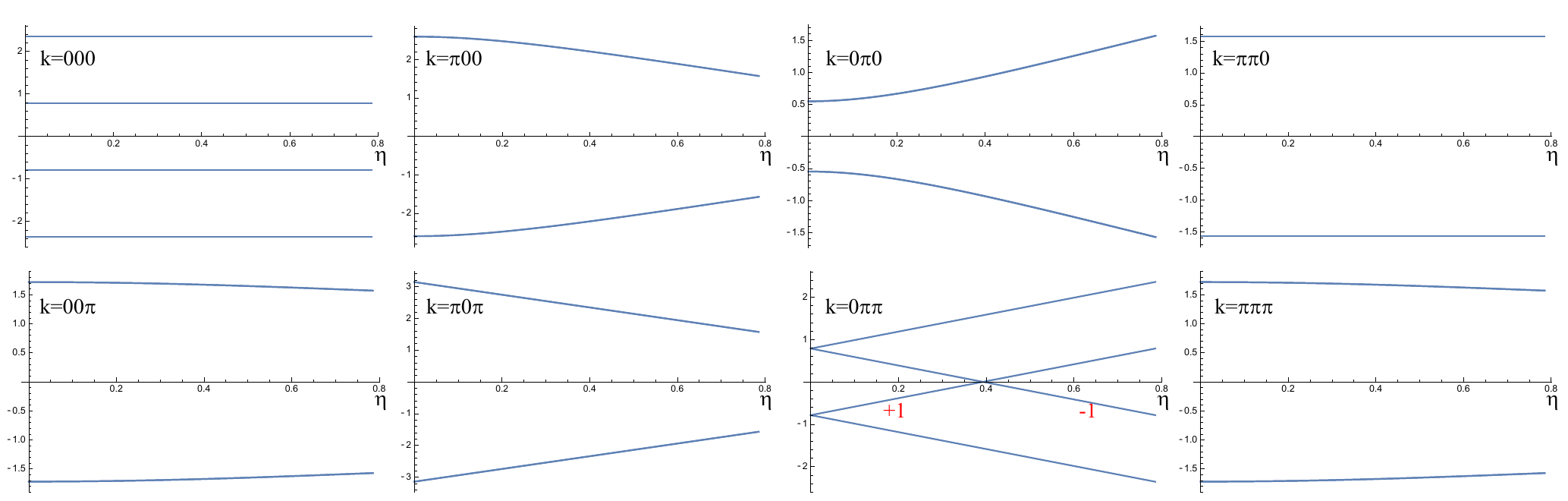}
\caption{The quasi-energies of the effective Hamiltonian $\mathcal{H}(\kk)$ at the eight inversion-invariant momenta as functions of $\eta$ for $0\le\eta<\frac{\pi}4$. The other parameters are set to $\mu=\delta=0$, $\gamma=\frac{\pi}8$.
The gap at $\kk=(0\pi\pi)$ closes at $\mu\approx 0.39$, where the red numbers represent the inversion eigenvalues.}
\label{fig:TBGap-e}
\end{figure}

As discussed in the second paragraph in this appendix, $U(\kk)$, as well as $\td{U}(\kk)$, is unitary.
We define the effective Hamiltonian of the Floquet system as the hermitian matrix
\begin{equation}
\mathcal{H}(\kk) = \frac{i}{4\Delta t} \ln \td{U}(\kk) .
\end{equation}
For convenience, we choose $\Delta t=1/4$ such that
\begin{equation}
\mathcal{H}(\kk) = i \ln \td{U}(\kk) .
\end{equation}
The eigenvalues of $\mathcal{H}(\kk)$ are referred to as quasi-energies, which range from $-\pi$ to $\pi$.
Since the Floquet system only has discrete time-translation symmetry, the quasi-energy is conserved mod $2\pi$.
For example, the $-\pi$ quasi-energy is same as the $\pi$ quasi-energy.
We now study the transitions as we tune $\mu$ from $-\frac{\pi}4$ to $\frac{\pi}4$.
Since the topological phase transition is always accompanied by gap closing, we assume that the low energy excitations of $\mathcal{H}$ correctly capture the topological phase transitions.
We take the parameters $\eta=\delta=0$, $\gamma=\frac{\pi}8$ and plot the quasi-energies at inversion-invariant-momenta as functions of $\mu$ for $-\frac{\pi}4<\mu<\frac{\pi}4$, as shown in \cref{fig:TBGap-M}.
As explained in the main text and \cref{app:limits}, the state at $\mu=-\frac{\pi}4$ is a trivial insulator and should have $z_{4}=0$, $z_{2,i}=0$ ($i=x,y,z$) due to the discussion in \cref{app:gap}.
As shown in \cref{fig:TBGap-M}, a gap closing between an even occupied state and an odd empty state happens at $\mu\approx -0.27$ and $\kk=(0\pi0)$; a second gap closing between an even occupied state and an odd empty state happens at $\mu\approx 0.27$ and $\kk=(0\pi0)$.
Due to \cref{eq:z2,eq:z4}, the intermediate phase ($-0.27<\mu<0.27$) has $z_4=1$, $z_{2,y}=1$, $z_{2,x}=z_{2,z}=0$ and hence is a WSM; the final phase ($\mu>0.27$) has $z_4=2$,  $z_{2,x}=z_{2,y}=z_{2,z}=0$ and hence is an axion insulator.

As discussed in the main text,  the parameter $\eta$ drives a transition to a 3D QAH state.
We take the parameters $\mu=\delta=0$, $\gamma=\frac{\pi}8$ and plot the quasi-energies at inversion-invariant-momenta as functions of $\eta$ for $0\le \eta <\frac{\pi}4$, as shown in \cref{fig:TBGap-e}.
As explained in the above paragraph, the state at $\eta=0$ is a WSM and has $z_4=1$, $z_{2,y}=1$, $z_{2,x}=z_{2,z}=0$.
As shown in \cref{fig:TBGap-e}, a gap closing between an even occupied state and an odd empty state happens at $\eta\approx 0.39$ and $\kk=(0\pi\pi)$.
Due to \cref{eq:z2,eq:z4}, the phase at $\eta>0.39$ has $z_4=2$, $z_{2,x}=0z_{2,y}=0$, $z_{2,z}=1$ and hence is a 3D QAH state with odd weak Chern number in the $z$-direction.

\begin{figure}[t]
\centering
\includegraphics[width=1\linewidth]{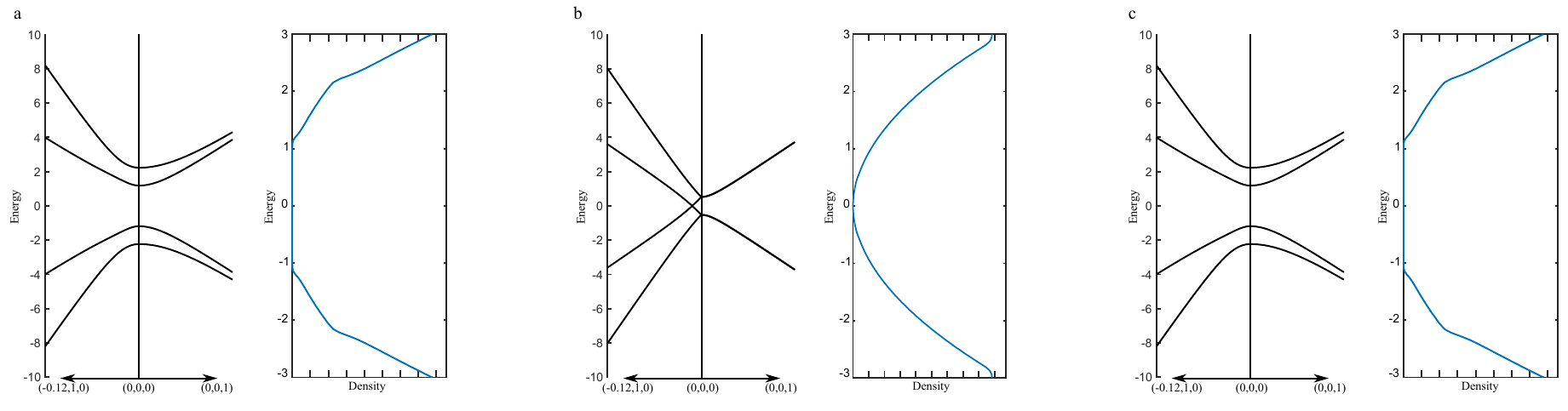}
\caption{The band structure and density of states of the k$\cdot$p Hamiltonian (\cref{eq:Hkp}).
(a) $\mu=-1$, trivial insulator.
(b) $\mu=0$, WSM.
(c) $\mu=1$, axion insulator.
}
\label{fig:TB-band}
\end{figure}

In order to confirm the trivial-WSM-axion transitions driven by $\mu$, we expand the effective Hamiltonian to linear order of $\kk$ around $\kk=(0,\pi,0)$ and $\mu$ at $\eta=\delta=0$.
We obtain the k$\cdot$p Hamiltonian
\begin{align}
\mathcal{H}(k_x,\pi+k_y,k_z) \approx  &
-\frac12(k_x-k_y) \sin^2\gamma \cdot \tau_0 \sigma_x
+ 2\mu \cos^2\gamma \cdot \tau_0 \sigma_y
-\frac12((k_x+k_y)\cos^2\gamma + k_z\sin^2\gamma ) \cdot \tau_0\sigma_z \nono\\
& + \frac14(k_x+k_y-2k_z)\sin 2\gamma \cdot \tau_x\sigma_x - \cos\gamma\sin\gamma \cdot \tau_x \sigma_y + \frac14(k_x-k_y) \sin 2\gamma \cdot \tau_x\sigma_z \nono \\
& - \cos\gamma\sin\gamma \cdot \tau_y\sigma_0 + \frac14( -2k_x + k_z + (2k_y -k_z )\cos2\gamma ) \cdot \tau_z\sigma_x + \sin^2\gamma \cdot \tau_z\sigma_y, \label{eq:Hkp}
\end{align}
where $\tau_{x,y,z}$, $\sigma_{x,y,z}$ are Pauli-matrices and $\tau_0$, $\sigma_0$ are identity matrices.
The inversion operator is given by \cref{eq:INV4}.
The quasi-energies and the corresponding parities ($\xi$) of $\mathcal{H}(0,\pi,0) $ can be analytically solved as
\begin{equation}
E_1= 2\mu \cos^2\gamma + \sin\gamma\sqrt{1+\cos^2\gamma},\qquad \xi_1=1,
\end{equation}
\begin{equation}
E_2= 2\mu \cos^2\gamma - \sin\gamma\sqrt{1+\cos^2\gamma},\qquad \xi_2=1,
\end{equation}
\begin{equation}
E_3= -2\mu \cos^2\gamma + \sin\gamma\sqrt{1+\cos^2\gamma},\qquad \xi_3=-1,
\end{equation}
\begin{equation}
E_4= -2\mu \cos^2\gamma - \sin\gamma\sqrt{1+\cos^2\gamma},\qquad \xi_4=-1.
\end{equation}
Applying \cref{eq:z4}, we obtain $z_4=0$ for $\mu<-\frac{\sin\gamma}{2\cos^2\gamma}\sqrt{1+\cos^2\gamma}$, $z_4=1$ for $-\frac{\sin\gamma}{2\cos^2\gamma}\sqrt{1+\cos^2\gamma}< \mu<\frac{\sin\gamma}{2\cos^2\gamma}\sqrt{1+\cos^2\gamma}$, and $z_4=2$ for $\mu>\frac{\sin\gamma}{2\cos^2\gamma}\sqrt{1+\cos^2\gamma}$.
For $\gamma=\frac{\pi}8$, $\mu=-1,0,1$ correspond to trivial insulator, WSM, and axion insulator, respectively.
The band structures and density of states for $\mu=-\frac{\pi}4,0,\frac{\pi}4$ are shown in \cref{fig:TB-band}a,b,c, respectively, where the WSM phase ($\mu=0$) has a gapless spectrum.

\begin{figure}[t]
\centering
\includegraphics[width=1\linewidth]{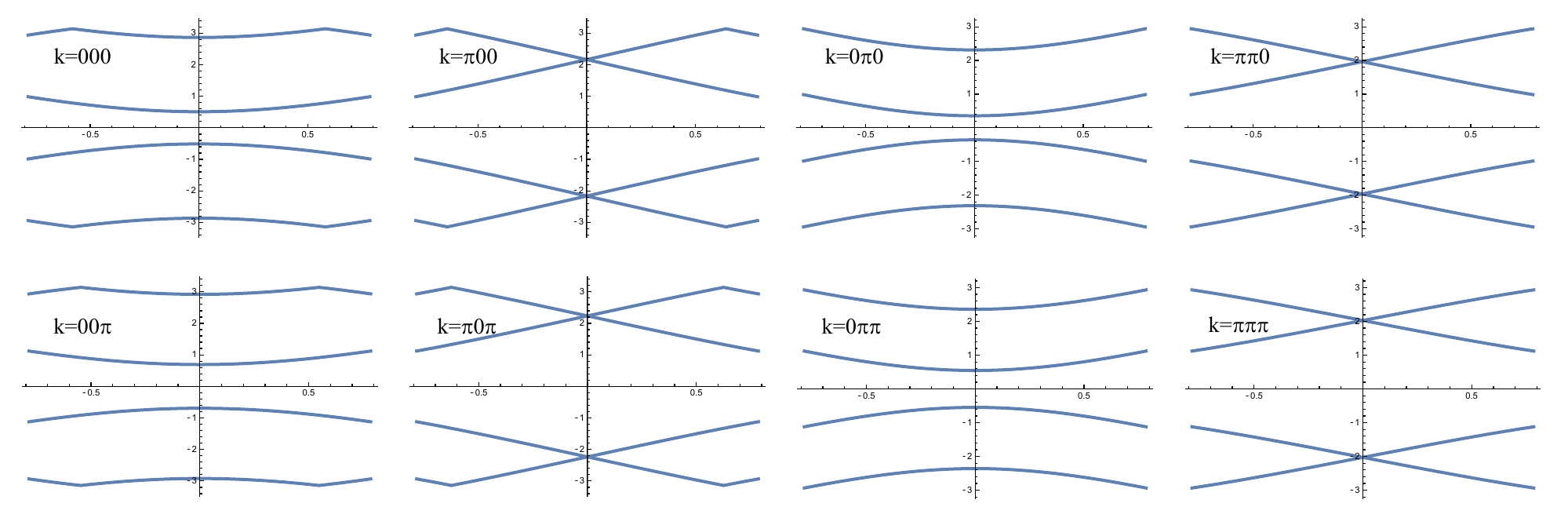}
\caption{The quasi-energies of the effective Hamiltonian $\mathcal{H}(\kk)$ at the eight inversion-invariant momenta as functions of $\mu$ for $-\frac{\pi}4<\mu<\frac{\pi}4$. The other parameters are set to $\eta=0$, $\delta=0.85$, $\gamma=\frac{\pi}8$.
With the inversion breaking term ($\delta$), we can see that the quasi-energies at all the high-symmetry momenta are all gapped for $-\frac{\pi}4<\mu<\frac{\pi}4$. 
}
\label{fig:Gap-NoInv}
\end{figure}

In the end, we study how the inversion breaking term ($\delta$) changes the gap closing process. In \cref{fig:Gap-NoInv}, we plotted the quasi-energies at the eight high symmetry momenta as a function of $\mu$ with $\delta=0.85$. 
We see that the gap closings at $\delta=0$ (\cref{fig:TBGap-M}) are now gapped. 
This implies the absence of delocalization transition from the trivial phase at $\mu=-\frac{\pi}4$ to the phase at $\mu=\frac{\pi}4$ that would have been axion insulator in the centrosymmetric case ($\delta=0$), under the condition that the inversion is strongly broken ($\delta=0.85$).

\end{document}